%% file: SADA.tex
\documentclass[lettersize,journal]{IEEEtran}
\usepackage{amsmath, amssymb, amsfonts}
\usepackage{algorithm}
\usepackage{array}
\usepackage[caption=false,font=normalsize,labelfont=sf,textfont=sf]{subfig}
\usepackage{textcomp}
\usepackage{stfloats}
\usepackage{url}
\usepackage{verbatim}
\usepackage{graphicx}
\usepackage{cite}
\usepackage{hyperref}

\usepackage{amssymb}% http://ctan.org/pkg/amssymb
\usepackage{pifont}% http://ctan.org/pkg/pifont
\newcommand{\cmark}{\ding{51}}%
\newcommand{\xmark}{\ding{55}}%

\usepackage{tikz}
\def\halfcheckmark{\tikz\draw[scale=0.4,fill=black](0,.2) -- (.15,0) -- (0.6,.5) -- (.15,.1) -- cycle (0.5,0.15) -- (0.53,0.15)  -- (0.4,0.5) -- cycle;}
\def\halfcmark{\textcolor{black}{\ding{51}}{\small\textcolor{black}{\kern-0.7em\ding{55}}}}

\DeclareMathOperator{\Hash}{H}
\DeclareMathOperator{\Hashuid}{\textup{H}_ {\textup{uid}}}
\DeclareMathOperator{\Hashcom}{\textup{H}_{\textup{com}}}
\DeclareMathOperator{\Hashagg}{\textup{H}_{\textup{agg}}}
\DeclareMathOperator{\Hashapp}{\textup{H}_{\textup{app}}}

\DeclareMathOperator{\Hashmask}{\textup{H}_{\textup{mask}}}
\DeclareMathOperator{\Hashvid}{\textup{H}_{\textup{vid}}}
\DeclareMathOperator{\E}{E}
\DeclareMathOperator{\HMAC}{HMAC}
\DeclareMathOperator{\Sign}{Sign}
\DeclareMathOperator{\Comm}{Commit}
\DeclareMathOperator{\Randassignl}{\xleftarrow{\$} \{0,1\}^{l_\text{hash}}}
\DeclareMathOperator{\Prb}{\text{Pr}}

\usepackage{algorithm}
\usepackage{algpseudocode}

\algnewcommand{\IIf}[1]{\State\algorithmicif\ #1\ \algorithmicthen}
\algnewcommand{\EndIIf}{\unskip\ \algorithmicend\ \algorithmicif}
\usepackage{tabularray}
\usepackage{booktabs}
\UseTblrLibrary{booktabs}
\SetTblrStyle{firstfoot,middlefoot,note-text}{\footnotesize\itshape}  
\usepackage{multirow} 
\usepackage[para,online,flushleft]{threeparttable}
\usepackage{array} %for centering with p commend

\usepackage{tabularx}
\makeatletter
\newcommand{\multiline}[1]{%
	\begin{tabularx}{\dimexpr\linewidth-\ALG@thistlm}[t]{@{}X@{}}
		#1
	\end{tabularx}
}
\makeatother

\usepackage{centernot}

\makeatletter
\newcounter{queryalg}
\newenvironment{queryalg}[1][htb]
{\renewcommand{\ALG@name}{Query}% Update algorithm name
	\let\c@algorithm\c@queryalg% Update algorithm counter
	\begin{algorithm}[#1]%
	}{\end{algorithm}}
\makeatother

\usepackage{amsthm}%for proof enviroment
\newtheorem{theorem}{Theorem}
\newtheorem{lemma}{Lemma}

\newtheorem{game}{Security Game}
\renewcommand\qedsymbol{$\blacksquare$}

\usepackage{dsfont} %mathds

\begin{document}

\title{Schnorr Approval-Based Secure and Privacy-Preserving IoV Data Aggregation}

\author{Rui Liu,~\IEEEmembership{Student Member,~IEEE}, and Jianping Pan,~\IEEEmembership{Fellow,~IEEE}
        % <-this % stops a space
\thanks{This work was supported in part by the Natural Sciences and Engineering Research Council (NSERC), in part by the Canada Foundation for Innovation (CFI), in part by the British Columbia Knowledge Development Fund (BCKDF) of Canada, and in part by the China Scholarship Council (CSC).}
\thanks{The authors are with University of Victoria, BC, Canada (e-mail: liuuvic@uvic.ca; pan@uvic.ca).}% <-this % stops a space
%\thanks{Manuscript received April 19, 2021; revised August 16, 2021.}
}



\maketitle

\begin{abstract}
Secure and privacy-preserving data aggregation in the Internet of Vehicles (IoV) continues to be a focal point of interest in both the industry and academia.
Aiming at tackling the challenges and solving the remaining limitations of existing works, this paper introduces a novel Schnorr approval-based IoV data aggregation framework based on a two-layered architecture. 
In this framework, a server can aggregate the IoV data from clusters without inferring the raw data, real identity and trajectories of vehicles. 
Notably, we avoid incorporating the widely-accepted techniques such as homomorphic encryption and digital pseudonym to avoid introducing high computation cost to vehicles. 
We propose a novel concept, data approval, based on the Schnorr signature scheme. With the approval, the fake data injection attack carried out by a cluster head can be defended against. The separation of liability is achieved as well.
The evaluation shows that the framework is secure and lightweight for vehicles in terms of the computation and communication costs. 
\end{abstract}

\begin{IEEEkeywords}
Vehicle clusters, privacy preserving, data aggregation, Internet of Vehicles, vehicular networks.
\end{IEEEkeywords}

\input{1_introduction}

\input{2_preliminaries}

\input{3_model}
\input{4_overview}
\input{4_details}
\input{5_evaluation}

\input{6_security}

\input{7_conclu}
\input{8_appendix}
\bibliographystyle{IEEEtran}  
\bibliography{article}

\begin{IEEEbiography}[{\includegraphics[width=1in,height=1.25in,clip,keepaspectratio]{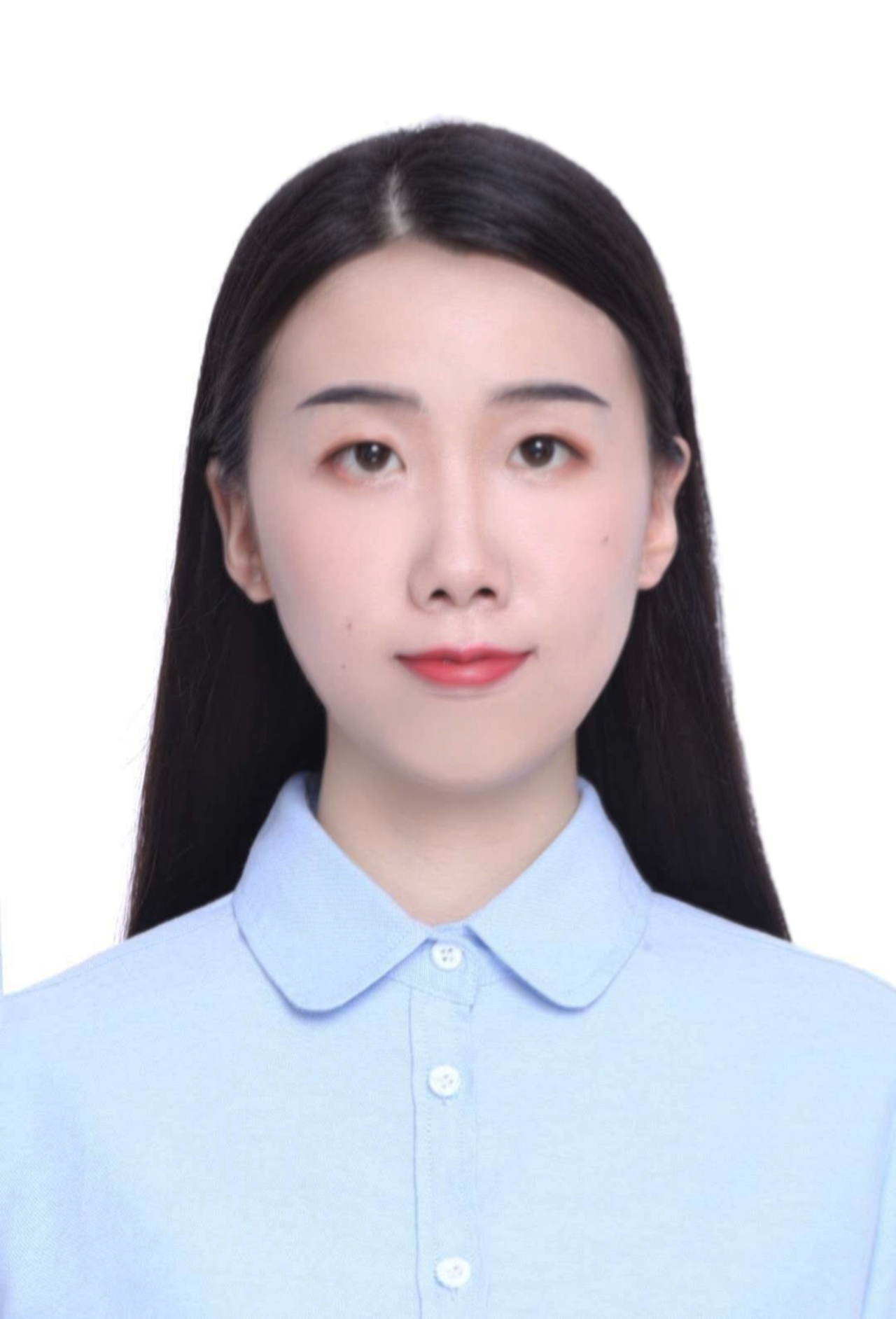}}]{Rui Liu}
	(Student Member, IEEE) received the Bachelor's degree in information security from China University of Geosciences, Wuhan, China, in 2018. She is currently a Ph.D. student under the supervision of Dr. Jianping Pan in the department of Computer Science, University of Victoria, Canada. 
	She received the National Scholarship (China) in 2016, the second prize in the Tenth National College Student Information Security Contest (China) in 2017, the University of Victoria Graduate Award from 2019 to 2023, the Best Poster Award for Workshop on Future Ubiquitous Networks (FUN) 2023-Spring (IEEE Victoria Section Joint VTS/ComSoc Chapter), the Gregory and Victoria Spievak Graduate Scholarship in 2023, and the IEEE Pacific Rim Wu-Sheng Lu Graduate Scholarship in 2023.  
	Her current research interests include security and privacy, Internet of Things, crowdsensing and federated learning.
\end{IEEEbiography}

\begin{IEEEbiography}[{\includegraphics[width=1in,height=1.25in,clip,keepaspectratio]{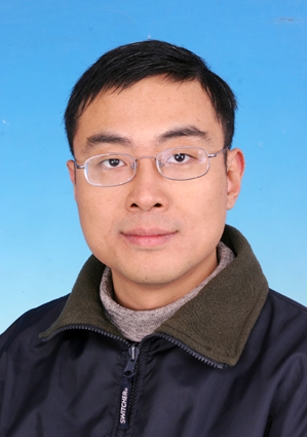}}]{Jianping Pan}
	(Fellow, IEEE) received the Bachelor's and PhD degrees in computer science from Southeast University, Nanjing, Jiangsu, China. He did his postdoctoral research at the University of Waterloo, Waterloo, Ontario, Canada. He also worked at Fujitsu Labs and NTT Labs. He is currently a professor of computer science at the University of Victoria, Victoria, British Columbia, Canada. He received the IEICE Best Paper Award in 2009, the Telecommunications Advancement Foundation's Telesys Award in 2010, the WCSP 2011 Best Paper Award, the IEEE Globecom 2011 Best Paper Award, the JSPS Invitation Fellowship in 2012, the IEEE ICC 2013 Best Paper Award, and the NSERC DAS Award in 2016, and has been serving on the technical program committees of major computer communications and networking conferences including IEEE INFOCOM, ICC, Globecom, WCNC and CCNC. He was the Ad Hoc and Sensor Networking Symposium Co-Chair of IEEE Globecom 2012 and an Associate Editor of IEEE Transactions on Vehicular Technology. His area of specialization is computer networks and distributed systems, and his current research interests include protocols for advanced networking, performance analysis of networked systems, and applied network security.
\end{IEEEbiography}

\end{document}

%% file: 1_introduction.tex
\section{Introduction}
\label{sec:intro}
% IoT data aggregation with VANET
\IEEEPARstart{I}n the Internet of Vehicles (IoV), vehicles equipped with various sensors and computing resources can provide many valuable data. It is not a brand new topic but gains ongoing interest in both industry and academia~\cite{yang2022privacy,cheng2023vfas}.
The sustained attention is driven by ongoing technological advancements, which give rise to many potential scenarios and applications.
For example, in smart city, traveling vehicles can monitor the air pollution or noise level at a fine granularity~\cite{wang2017efficient, liu2023lightweight}. With federated learning, vehicles can perform as local workers, contributing their local training results to build a globally shared learning model for parking space estimation~\cite{huang2021fedparking}. %The energy consumption value can be gathered for speed advisory of vehicle platoons~\cite{li2023eco}.
The speed of vehicles, directions, and road conditions are also  invaluable for traffic management and autonomous driving~\cite{9229189,10091215}.

% problems: security of the data and privacy
%Protecting the sensory data, and the identities and trajectories of vehicles is important.
One typical scenario is that a central server aggregates all the sensory data from individual vehicles, where protecting the raw data, and the identities and trajectories of vehicles is important.
Although many solutions have been proposed, researchers never stop their efforts to enhance the security, efficiency, and practicability.
Kong~{\em et al.}~\cite{kong2019privacy} present a multi-dimensional IoV data collection scheme. The modified Paillier cryptosystem is adopted to protect the sensory data and location information maintained in the reports, whereas the real identity of vehicles is exposed to the server. An adversary can further infer the trajectories of vehicles by linking the communication activities at different road segments. 
Li~{\em et al.}~\cite{li2023eco} propose a safe and eco-friendly speed advisory system. Each vehicle encrypts the sensory data with an ElGamal commitment. However, to get the aggregation of the data, a discrete logarithm problem should be solved, which is time-consuming. To preserve the privacy, pseudonyms are used. Additional work is required for pseudonym management.

In the previous work~\cite{liu2023crs} (named as \textbf{CRS}), we propose a two-layered architecture for privacy-preserving data aggregation in federated learning.
In CRS, a cloud server would like to aggregate sensory data from the vehicles via road-side units (RSUs). 
Different from traditional end-to-end architecture, most of the vehicles do not need to directly upload data to the server (i.e., do not need to directly communicate with RSUs): clusters are formed with multiple vehicles, and a cluster head is elected for each cluster to locally aggregate the data and perform as a gateway between the cluster and RSUs. As a result, both the communication cost of vehicles and risks of exposing data, vehicle identities and vehicle trajectories to RSUs and the server are reduced. 
To further preserve the data privacy, the technique of data masking is adopted, with which the cluster head can locally calculate the average of all the sensory data without acquiring the raw data. 
However, a \textit{bad} cluster head may upload a fake average value. A newly emerged challenge in the two-layered architecture is ensuring the authenticity of the average. 
Besides, a digital signature-based anonymous authentication scheme is used in CRS to preserve the identity and trajectory privacy of vehicles. The pseudonym generation, distribution and management involve high overhead. 

To tackle the above challenges while avoiding adopting the typical techniques which may bring high computation and communication costs to vehicles, in this paper, we propose a novel \textbf{S}chnorr \textbf{A}pproval-based IoV \textbf{D}ata \textbf{A}ggregation framework, \textbf{SADA}, over the two-layered architecture. 
In SADA, a cluster head securely aggregates the sensory data in a cluster with a regional data collection scheme, named \textbf{CluCol} (\textbf{Clu}ster \textbf{Col}lection). It relays the locally aggregated result to a cloud server via RSUs anonymously with a novel Schnorr signature-based approval. 
To be specific, the main work and contributions of the paper are as follows:
\begin{itemize}
	\item Instead of adopting the original masking technique as CRS, we propose a recoverable masking technique to preserve the privacy of sensory data. A significant advantage is that the technique enables a cluster to recover from an input invalidation attack. Different from differential privacy, it does not reduce the data accuracy. It is more lightweight for vehicles than the encryption-based solutions. 
	\item We present a new concept, \textit{data approval}, based on multi-signature schemes. It addresses the major limitation of CRS: it can prevent a cluster head from uploading fake aggregated data on behalf of the whole cluster. Invalid approvals can be detected fast with a binary tree-based pre-checking algorithm. We formally prove the security of the approval scheme in the Random Oracle Model (ROM).
	\item The identity and trajectory privacy of the cluster head is preserved with the cluster approval. Individual vehicles are hidden behind the cluster head. Thus, there is no need for updating or management of pseudonyms, which saves the computation and communication resources of vehicles.
	\item Compared with CRS, SADA provides enhanced security protection and circumvents the substantial high overhead of pseudonym techniques. It also achieves the separation of liability among vehicles, which is a valuable advantage in the two-layered architecture. A misbehaved cluster head can be identified with a cluster key audit strategy. 
\end{itemize}

In the rest of the paper, we introduce the related work and cryptography tools in Sections~\ref{subsec:rela} and~\ref{subsec:tools}, respectively. 
We briefly introduce how we modify or use them differently.
A system model and threat model are provided in Section~\ref{sec:model}.
In Section~\ref{sec:overview}, we briefly introduce the framework design, followed by a detailed description in Section~\ref{sec:details}.
In Section~\ref{sec:perfor}, we analyze and evaluate the performance of the work. Simulations have been conducted in ns-3 (Network Simulator-3)~\cite{ns3}. 
In Section~\ref{sec:sec}, we compare, analyze and prove the security of SADA. A conclusion is given in Section~\ref{sec:con}.

%% file: 2_preliminaries.tex
\section{Related Work}
\label{subsec:rela}
To preserve the privacy of the sensory data,
one typical technique adopted in IoT is differential privacy where noise is added to the raw data. 
Raja~{\em et al.}~\cite{raja2020sp} propose a secure and private-collaborative intrusion detection system, SP-CIDS, for VANETs. To train a detection model in a distributed and privacy-preserving manner, vehicles perturb the local training results with Laplace noise.
%The system allows a vehicle to detect malicious events from network audit data with a machine learning model. To train the model in a distributed and privacy-preserving manner, vehicles exchange their local training result which are perturbed with Laplace noise. 
However, the accuracy of the aggregated value is affected. 
Similarly, Tang~{\em et al.}~\cite{tang2019secure} propose a health data aggregation scheme where each health center adds Laplace noise to the data.
To further protect the perturbed data, the Boneh–Goh–Nissim (BGN) cryptosystem is adopted, which is additively homomorphic. 
The server can obtain the sum of the data from all health centers without decrypting the individual ciphertexts.
Kong~{\em et al.}~\cite{kong2019privacy} adopt the modified Paillier cryptosystem for privacy-preserving sensory data collection in VANETs, which also has the homomorphic property. 
Li~{\em et al.}~\cite{li2021lightweight} improve the labeled homomorphic encryption (LabHE) cryptosystem to tackle the privacy challenge among the data owners, untrustworthy servers, and the data users in the Industrial Internet of Things (IIoT). 
Although these homomorphic encryption algorithms are widely adopted because of the usability and precision of the aggregation result, the computation and communication costs are relatively high. 

Bonawitz~{\em et al.}~\cite{bonawitz2017practical} propose a pairwise masking technique for federated learning. It allows a server to compute the sums of model parameter updates from individual clients in a secure and efficient way. To handle the dropped-out clients, a threshold secret sharing scheme is adopted for mask recovery. Compared with differential privacy, this technique can maintain the precision of the data. Compared with homomorphic encryption, it is efficient in computing.
Thus, we adopt the masking technique in CRS~\cite{liu2023crs} and propose a secure and reliable data gathering protocol based on network coding. In the current work, SADA, we still adopt this technique for secure data aggregation but improve it for bad vehicle exclusion. The introduction of the technique, its limitations, and our corresponding improvements can be found in Sections~\ref{subsec:tools} and~\ref{subsec:masking}.

To preserve the identity and trajectory privacy of vehicles, pseudonym is a widely-accepted technique. 
Lu~{\em et al.}~\cite{wei2020secure} propose a privacy-preserving authentication scheme for emergency traffic message transmission in VANETs. Each vehicle generates a pseudo identity with an unhackable tamper-proof device. 
In Eco-CASA~\cite{li2023eco}, vehicles are allowed to register multiple anonymous key pairs but the authors do not provide details.
In CRS, we adopt the work of Moni~{\em et al.}~\cite{moni2021scalable}. In this work~\cite{moni2021scalable}, the authors propose a secure authentication and message verification scheme for vehicular ad-hoc networks (VANETs). Regional trusted authorities issue pseudo-IDs to vehicles, which are used in all communications to preserve the privacy. 

In CRS~\cite{liu2023crs}, to protect the identities and trajectories of vehicles from RSUs and the server, we propose the two-layered architecture. Member vehicles are hidden behind cluster heads so that are protected. As for cluster heads, we adopt the pseudo-ID-based solution~\cite{moni2021scalable} for anonymous communication.
A limitation is that the pseudonym technique requires additional work in the management and updating of the pseudo-IDs and cryptographic keys. 
Thus, in this work, first, we adopt the architecture in CRS~\cite{liu2023crs} to protect the member vehicles in clusters. Then, we try to preserve the privacy of cluster heads in a different way: data approval.

\section{Preliminaries}
%\label{sec:pre}
%\subsection{Cryptography tools and adopted techniques}
\label{subsec:tools}

In this section, we introduce the basic cryptography tools and techniques used in the proposed framework. 
We also briefly discuss the necessity and rationale for making revisions on them and the corresponding differences.
%masking+shamir--> one section
%multi-->discuss more on difference
\subsection{Masking Technique}
Data masking allows a server to aggregate data from client parties in a secure way: each pair of clients $(u_i, u_j)$ agrees on a mask $\alpha_{i,j}$ with the Diffie-Hellman protocol. Each client can generate a masked value for its secrets with all the masks it obtains. The server can calculate the sum of all the secrets only with the masked values. Details can be found in~\cite{bonawitz2017practical, liu2023crs}.

To handle dropped clients during the sum calculation, Bonawitz~{\em et al.}~\cite{bonawitz2017practical} adopt a threshold secret sharing scheme. Each Diffie-Hellman secret used for generating masks is divided into multiple shares, which are then distributed to different clients. The masks can be recalculated with a threshold number of shares.
As a result, if a client drops before providing its masked value, the corresponding masks can be removed, and the sum of all the secrets can still be correctly calculated. 
Although we do not target handling packet loss or client drop as what we achieved in CRS, we adopt the masking technique with Shamir's secret sharing to preserve data privacy and enable recovery from an input invalidation attack.
%we are inspired by this work. 
%To preserve the data privacy and recover from an input invalidation attack, we adopt the data masking technique with Shamir's secret sharing. 
Different from the original design~\cite{bonawitz2017practical}, there is no need to recalculate masks, which saves the computation cost. 
%Details are further given in Alg.~\ref{alg:alg_mask}. 

\subsection{Schnorr Signatures}
The Schnorr signature scheme is a simple and efficient digital signature scheme. 
The key-prefixed variant works as follows: the signer and verifier agree on a cyclic group $\mathbb{G}$ of prime order $q$ with generator $g$. The public key of the signer is $y=g^x$ where $x\in \mathbb{Z}_q^{\times}$ is the private key. 
%Given a message $m$, the signature is $(s =k+xe,R=g^k)$, where $k\xleftarrow{\S}  \mathbb{Z}_q^{\times}$ ($\xleftarrow{\S} $ denotes randomly sample a value from a group). $e=\Hash(y\|R\|m)$ where $\Hash$ is a hash function.
Given a message $m$, the signer selects $k\xleftarrow{\S}  \mathbb{Z}_q^{\times}$ ($\xleftarrow{\S} $ denotes randomly sample a value from a group), and calculates $R=g^k$, $e=\Hash(y\|R\|m)$ and $s =k+xe$, where $\Hash$ is a hash function. The pair $(s,R)$ is the signature on $m$.
The verifier calculates $e^\prime=\Hash(y\|R\|m)$ and checks $g^s \stackrel{?}{=} Ry^{e^\prime}$. 

An advantage of the scheme is that it supports batch verification, which is faster than individually verifying each one.
%: a large number of signatures can be verified together, which is faster than individually verifying each one.
%With the Schnorr signature scheme, a large number of signatures can be verified together, which is faster than individually verifying each one.
Suppose there are $n_{\text{m}}$ signers. Each signer with an index $i$ generates a signature $(s_i,R_i)$ with its public key $y_i$ on its message $m_i$. The verifier can verify all the signatures together by checking $\prod_{i=1}^{n_{\text{m}}} g^{s_i  a^1_i} \stackrel{?}{=} \prod_{i=1}^{n_{\text{m}}} \{R_iy_i^{e_i^\prime}\} ^{a^1_i}$, where $a^1_i$ are random numbers and $e_i^\prime=\Hash(y_i\|R_i\|m_i)$. 
%The Bos-Coster's algorithm can be used to improve the calculation efficiency.
The equation can be calculated faster by some mathematical computation algorithms (e.g., the Bos-Coster's algorithm). 

\subsection{Multi-Signature}
A Schnorr multi-signature scheme allows a group of signers to produce a joint signature on a common message. The size of the joint signature is exactly the same as a regular Schnorr signature. With an aggregated public key, a verifier can verify the joint signature as normal. There is no need to expose the individual keys.
The naive scheme is vulnerable to the rogue-key attack, where the aggregator can generate a joint signature on behalf of all the signers by itself. 
In our work, we adopt a new rogue-key attack-resistant variant, MuSig~\cite{maxwell2019simple}.
However, it cannot directly satisfy the specific security and efficiency requirements in the two-layered architecture. Thus, we propose a new concept, data approval, and use MuSig as the foundation of it. We also call it \textit{Schnorr approval}.
In Section~\ref{subsec:data_approval}, we explain the rationale and the minimal but intelligent modifications we made on MuSig.

%% file: 3_model.tex
\section{System and Threat Model}
\label{sec:model}
In this section, we give the detailed system model and threat model. The frequently used notations are summarized in Table~\ref{4:tab:notations} for reference.

\begin{table}[!t]
	%\vspace{-3mm}
	%\renewcommand{\arraystretch}{1.3}
	\centering
	\small
	\caption{Frequently Used Notations}
	\label{4:tab:notations}
	\begin{tblr}{colspec={X[1,l]|X[4.5,l]}}
		\toprule
		Notation & Description \\
		\midrule
		%\hline
		$(\mathit{skr},\mathit{pkr})$ & The private and public key pair of an RSU
		\\
		$\{v_i \mid i \in N_\text{v}\}$ & A cluster composed with $n$ vehicles ($N_\text{v}=\{1,2,\dots, n_\text{v}\}$)
		\\
		$\mathit{CH}$ & A cluster head
		\\
		$(\mathit{sk}_i,\mathit{pk}_i)$ & The private and public key pair of $v_i$
		\\
		$\mathit{CS}$ & The cloud server
		\\
		$(\mathit{sks},\mathit{pks})$ & The private and public key pair of $\mathit{CS}$
		\\
		$(\mathit{ska},\mathit{pka})$ & The private and public key pair of the TA
		\\
		$\mathit{UID}$ & ID of a data aggregation event
		\\
		$\alpha_{i,j}$ & The mask agreed between $v_i$ and $v_j$
		\\
		$\mathit{data}_i$, $c_i$ & The individual sensory data and the corresponding masked data of $v_i$
		\\
		$\overline{\mathit{data}}$& The aggregated value (i.e., the average of all $\mathit{data}_i$) of a cluster
		\\
		$v_{i_{\text{re}}}$ & An adversarial member vehicle (index $i_{\text{re}}$)
		\\
		$\beta_i$ & The reconstruction parameter generated by $v_i$
		\\
		$\mathit{appr}_i$ & The sub-approval of $v_i$
		\\
		$\mathit{appr}$ & The cluster approval
		\\
		$\widetilde{\mathit{pk}}$ & The aggregated public key of a cluster
		\\
		$\mathit{cre}_{i_\text{CH}} $ & The credential of $\mathit{CH}$ with an index $i_\text{CH}$
		\\
		$\mathit{key}_1$, $\mathit{key}_2$ & Two session keys between $\mathit{CH}$ and the RSU
		\\
		$L_\text{rc}$ & A list of records from all member vehicles
		\\
		\bottomrule
	\end{tblr}
\end{table}
%	\vspace{-3mm}
\subsection{System Model}
\label{subsec:sysmodel}
\begin{figure}[htb!]
	\centering
	\includegraphics[width=1.0\linewidth]{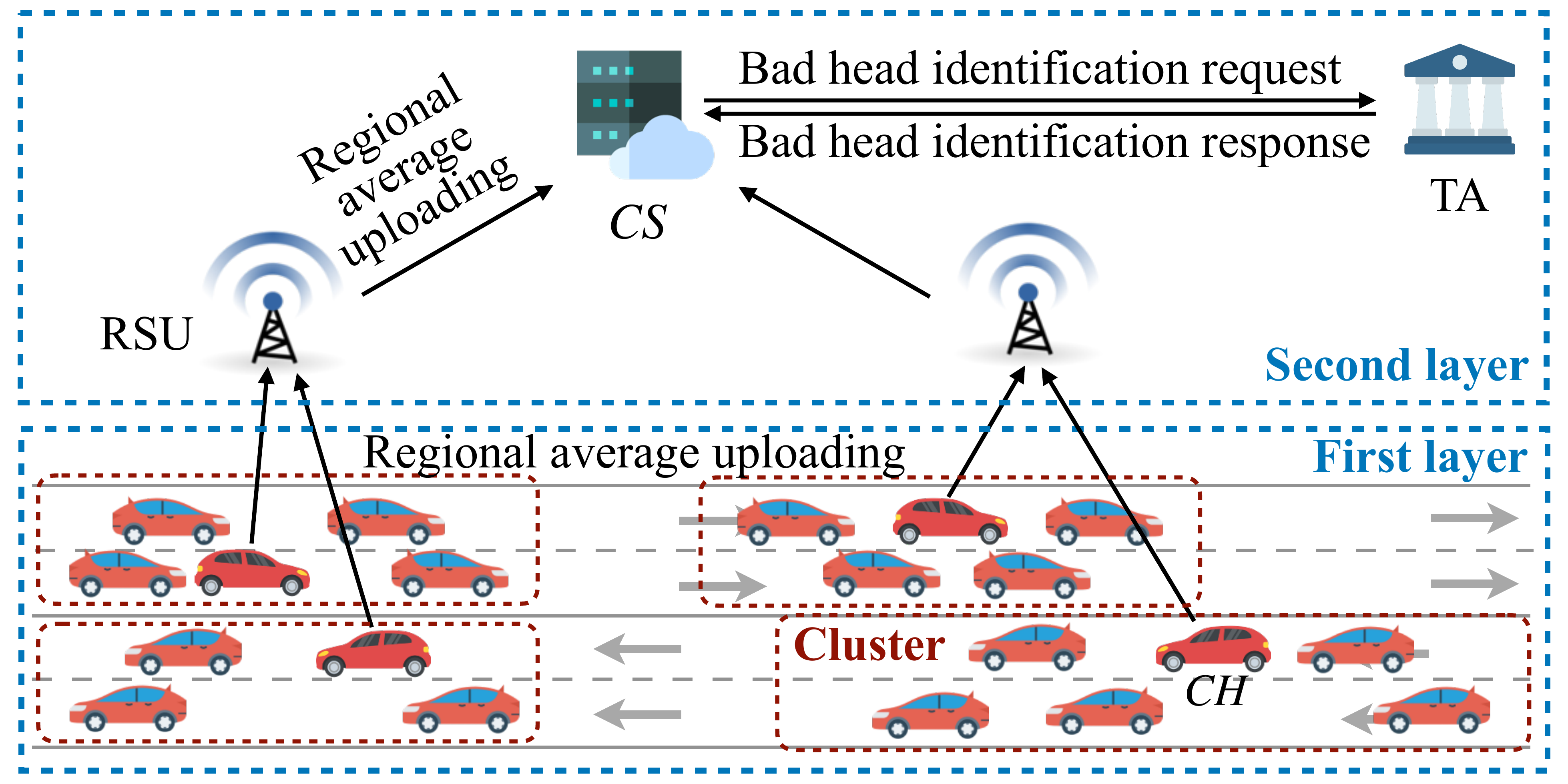}
	\caption{Framework model.} 
	%\vspace{-1mm}
	\label{fig:frame}
\end{figure}

As shown in Fig.~\ref{fig:frame}, we formally define the entities:
\begin{itemize}
	\item \textbf{Road-side units}: RSUs are base stations on the roadside. It relays the messages from vehicles to the server by wireless communication.
	The private and public key pair of an RSU is $(\mathit{skr}, \mathit{pkr})$. 
	\item \textbf{Vehicle clusters}: a bunch of $n_{\text{v}}$ vehicles traveling in the same direction, $\{v_i \mid i \in N_{\text{v}}\}$, compose a \textit{cluster}. $N_{\text{v}}=\{1,2,\dots, n_{\text{v}}\}$.
	Every vehicle keeps a sensory data, denoted by $\mathit{data}_i$. A \textbf{cluster head} $\mathit{CH}$ performing as a gateway between the cluster and an RSU has the following tasks: a) to acquire data and verify the messages from member vehicles and calculate an average of the data; b) to pre-check the data approval; c) to identify misbehaved member vehicles, and generate a new average and a new approval if needed; d) to send the average, approval and necessary parameters to RSUs. All the other vehicles are called \textbf{cluster members}. The main tasks of a member vehicle in each \textit{sensing cycle} (i.e., a constant time slot) include generating a sub-approval, sending it and $\mathit{data}_i$ to $\mathit{CH}$, and helping $\mathit{CH}$ in new average calculation when needed. The key pair of a member vehicle $v_i$ is denoted by $(\mathit{sk}_i,\mathit{pk}_i)$. 
	\item \textbf{Cloud server}: the server $\mathit{CS}$ (with a key pair $(\mathit{sks}, \mathit{pks})$) runs in the cloud computing environment.
	The aggregated data are processed by $\mathit{CS}$ in each \textit{handling cycle}. $\mathit{CS}$ also works on identifying misbehaved cluster heads with the help of a trusted authority. 
	\item \textbf{Trusted authority (TA)}: the TA (with a key pair $(\mathit{ska},\mathit{pka})$) generates, distributes and manages the identities, credentials and cryptographic keys for all legal entities in the system. It has the responsibility to respond to bad head identification requests from $\mathit{CS}$.
	%The key pair of the TA is $(\mathit{ska},\mathit{pka})$. 
\end{itemize}

In our previous work CRS, we consider how to handle packet loss and discuss the reliability of clusters and the related state-of-the-art research~\cite{liu2023crs}. Thus, in this work, we assume the communications are reliable. 
All packets can be sent and received successfully in the network.  
%We assume a cluster is stable in a sensing cycle. 
%The member join, the member departure and the head update are not considered but further discussed in Section~\ref{sec:perfor}. 

\subsection{Threat Model}
\label{subsec:threatmodel}
The entities who are not registered in the system are considered as external adversaries. They have the ability to capture messages transmitted between two parties. 
As for the registered entities, the assumptions are as follows:
\begin{itemize}
	\item We assume the TA is fully trusted. 
	\item RSUs and $\mathit{CS}$ are honest-but-curious: they firmly follow the protocol but may be curious about the sensory data, vehicle identities and vehicle trajectories. Besides, to infer the vehicle trajectories, RSUs may collude with each other and link the communication activities of a vehicle of interest.
	\item Vehicles are considered as partial-honest-and-curious. All the vehicles are curious about the sensory data of others. 
	In the partial-honest model, all vehicles are honest on their own individual sensory data, but may engage in specific adversarial behaviors:
	a) a \textit{bad} member vehicle may provide an invalid sub-approval to $\mathit{CH}$ which can lead to the invalidation of the aggregated data;
	b) a \textit{bad} $\mathit{CH}$ may upload a fake regional average, i.e., the cluster average of all the individual data.
	The assumptions are reasonable because the two roles (i.e., a member vehicle or $\mathit{CH}$) hold different rights and impacts: the average uploaded by $\mathit{CH}$ representing all vehicles in the cluster has a higher impact than the individual data of one vehicle. 
	%	A vehicle may change its role (i.e., a member vehicle or $\mathit{CH}$) in different sensing cycles. Considering the differences in the rights and impacts of the two roles, 
\end{itemize}

We summarize the major attacks we target within the threat model:
\begin{itemize}
	\item \textbf{Data leakage attack}: the sensory data may be exposed to vehicles, RSUs, $\mathit{CS}$ and external adversaries.
	\item \textbf{Identity leakage attack}: the real identities of vehicles may be exposed to RSUs,  $\mathit{CS}$ and external adversaries.
	\item \textbf{Trajectory tracking attack}: the trajectories of vehicles may be inferred by RSUs,  $\mathit{CS}$ and external adversaries.
	\item \textbf{Fake data injection attack}: a bad $\mathit{CH}$ may try to upload a fake average value or fake approval to the RSU.
	\item \textbf{Input invalidation attack}: %an internal adversary (i.e., a \textit{bad} member vehicle) 
	a bad member vehicle may provide an invalid sub-approval. It may also shift the accountability and potential punishment from the adversary to $\mathit{CH}$.
	\item \textbf{Eavesdropping attack}: the messages sent in the system may be captured by an adversary.
	\item \textbf{Message forgery and tempering attacks}: the messages may be forged or tempered by an adversary.
\end{itemize}

In addition, we aim at tracing the bad vehicles and removing the negative impact of an attack as much as possible: a) $\mathit{CS}$ should be able to not only detect the fake data injection attack, but also trace the bad $\mathit{CH}$ who anonymously uploaded the fake average; b) when an input invalidation attack is detected,  
$\mathit{CH}$ should be able to identify the bad member vehicle and exclude it from the protocol execution. A new regional average value without the sensory data of the bad vehicle should be calculated.

%% file: 4_overview.tex
\section{Framework Overview}
\label{sec:overview}
\begin{figure}[htb!]
	\centering
	\includegraphics[width=1\linewidth]{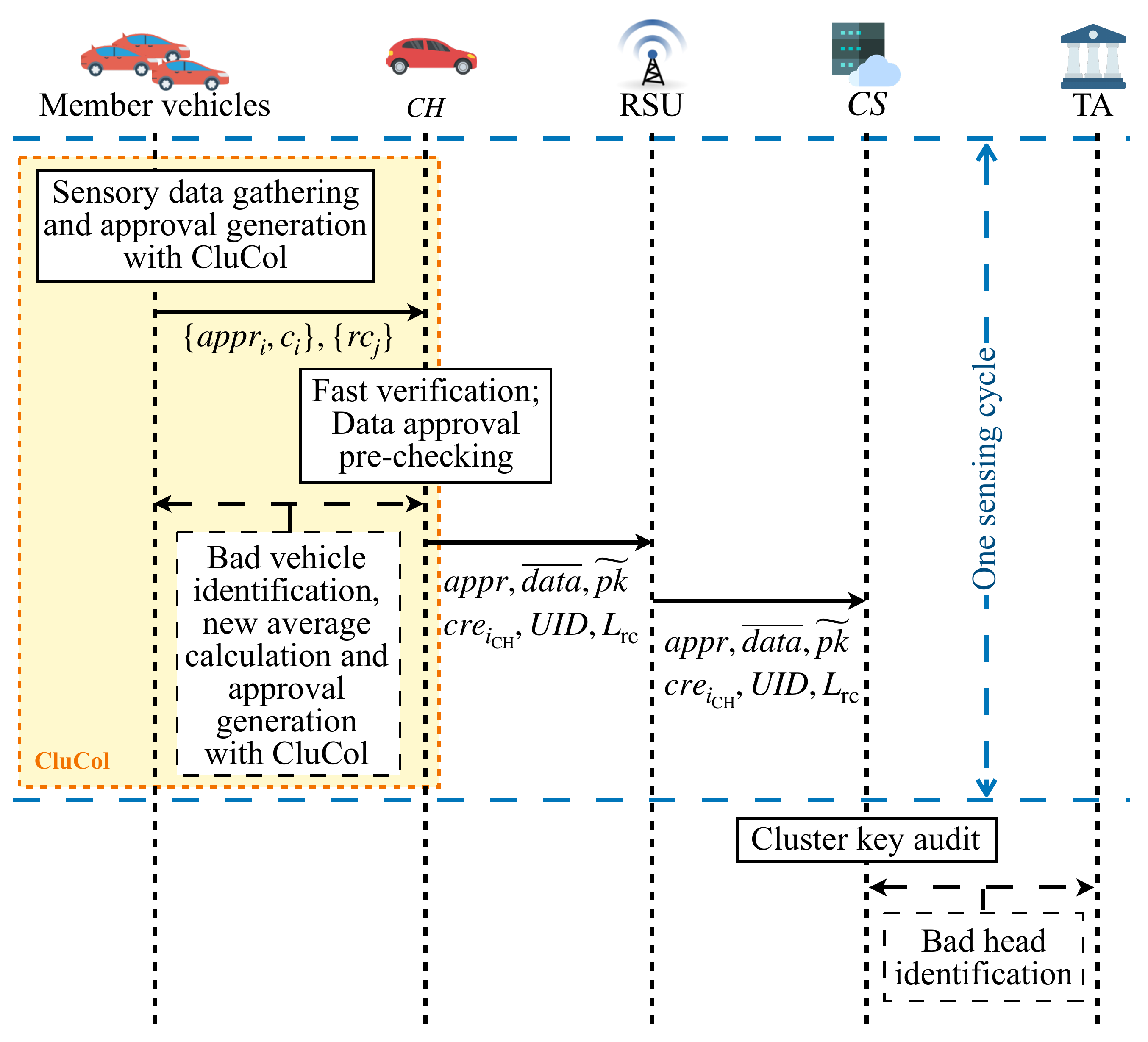}
	\caption{Protocol of SADA.} 
	\label{fig:pro}
	%\vspace{-4mm}
\end{figure}

The main protocol of the two-layered privacy-preserving data aggregation framework, \textbf{SADA}, is shown in Fig.~\ref{fig:pro}. In this section, we briefly introduce the main idea. The notations, messages, algorithms and detailed protocols are introduced in Section~\ref{sec:details}. 

The first layer of SADA is composed of vehicle clusters. 
All vehicles in the cluster can generate the same ID with a hash function $\Hashuid$,  a pre-agreed timestamp $\mathit{tmp}^1$ and a uniquely encoded list of public keys of the vehicles $L_{\text{pk}}$:
%\footnote{We assume $L_{\text{pk}}$ is uniquely encoded first, i.e., using a fixed order. }:
\begin{equation}
\mathit{UID} = \Hashuid(L_{\text{pk}} =\{\mathit{pk}_i\mid i\in N_{\text{v}}\}\| \mathit{tmp}^1).
\end{equation} 
With binding $\mathit{tmp}^1$ and the information of all vehicles, we can use $\mathit{UID}$ to uniquely identify a data aggregation event, i.e., collecting data in a specific sensing cycle within a specific cluster.

We assume $\mathit{CH}$ is first elected by all the vehicles in the cluster\footnote{We omit the details of head election because it has been explored in our previous work~\cite{liu2023crs}.}.
%, which has been explored in our previous work~\cite{liu2023crs}.
%\footnote{We omit the details of head election because it has been explored in our previous work, CRS.}.
In this process, a symmetric key $\mathit{key}_{\text{v}_{i,j}}$ and a mask $\alpha_{i,j}$ are exchanged and agreed upon by each vehicle pair $(v_i, v_j)$ with the Diffie–Hellman key exchange protocol, as a preparation for regional data collection.
The proposed lightweight and secure regional data collection scheme \textbf{CluCol} is specifically designed for vehicle clusters. 
It is the core component of SADA. 

The design objective of CluCol is to allow $\mathit{CH}$ to securely obtain $\overline{\mathit{data}}$, the average of all the data provided by every member vehicle in the cluster. 
%It is necessary to exchange or provide $\mathit{data}_i$ to other vehicles or $\mathit{CH}$.
To guarantee the confidentiality of $\mathit{data}_i$, $v_i$ first masks it as $c_i$ with the exchanged masks. An accurate sum can be calculated without acquiring any mask or $\mathit{data}_i$. This is the design in CRS. In SADA, the difference is that, instead of adopting the original masking technique, we propose a \textbf{recoverable masking protocol} integrating the masking technique with Shamir's secret sharing. When a bad member vehicle is identified, a new $\overline{\mathit{data}}$ can be easily calculated with the help of $t_{\text{sm}}$ (a threshold) good member vehicles. 
There is no need to restart the process from the beginning. This design makes it possible for a cluster to recover from an input invalidation attack. We wish to emphasize that the goal and design of the integration of secret sharing is different from that in~\cite{bonawitz2017practical}. Details are given in Section~\ref{subsec:masking}.

In CRS, to aggregate the regional data, $\mathit{CH}$ collects $\{c_i \mid i\in N_{\text{v}}\}$ and calculates the average $\overline{\mathit{data}}$ directly. Recall that we try to mitigate the limitation in CRS and assure the authenticity of $\overline{\mathit{data}}$. An intuitive design is to use the Schnorr multi-signature scheme, where each vehicle signs on $\overline{\mathit{data}}$ and $\mathit{CH}$ calculates the joint signature. 
However, this design cannot meet all the requirements and defend against all the targeted attacks. $\mathit{CH}$ may send a fake $\overline{\mathit{data}}$ to all member vehicles and obtain a valid joint signature. To defend against this fake data injection attack, we try to propose a new concept, approval, by adjusting a Schnorr multi-signature scheme, MuSig. 

The approval involves the contribution of all $v_i$ and confirms the authenticity of $\overline{\mathit{data}}$. In the proposed \textbf{data approval generation protocol}, different from the intuitive design, $\overline{\mathit{data}}$ is calculated by each $v_i$ independently for authenticity. 
The commitments of $\{c_j \mid j\in N_{\text{v}} \wedge j\neq i\}$ are collected before sending $c_i$ to other vehicles. 
%It is achieved by adjusting the protocol of MuSig. 
Each vehicle generates a \textit{sub-approval} $\mathit{appr}_i$ for $\overline{\mathit{data}}$.
After collecting $\mathit{appr}_i$ from all member vehicles, $\mathit{CH}$ generates the cluster approval $\mathit{appr}$. The details are given in Section~\ref{subsec:data_approval}. 

We want to emphasize the necessity and rationale behind introducing the new concept. In a traditional multi-signature scheme, the joint signature is produced on a predetermined shared message, usually a transaction. Without all the required parties (i.e., signers), the transaction cannot be authorized and completed. Differently, with the proposed approval scheme, there is no transaction or a pre-agreed common message. All parties (i.e., vehicles) are responsible for calculating an average locally. The primary objective is to approve that the final average value is indeed calculated based on each party's individual data. It is specifically designed for the scenario of data aggregation within a vehicle cluster. 

Another attack we target is the input invalidation attack. To defend against it, we present a \textbf{data approval pre-checking algorithm} in Section~\ref{subsec:appr_check}. With a thoughtful design, $\mathit{CH}$ can check the validation of $\mathit{appr}$ and efficiently trace the bad sub-approvals by maintaining intermediate results in a tree structure. 
The bad member vehicles can be excluded from the protocol execution with the proposed recoverable masking protocol.

%With the proposed recoverable masking protocol, a new $\overline{\mathit{data}}$ can be calculated with the help of $t_{\text{sm}}$ good member vehicles where $t_{\text{sm}}$ is a threshold. 

To provide secure communication and fast verification in a cluster, the messages sent from member vehicles to $\mathit{CH}$ are signed with the Schnorr signature scheme, which can be verified in batch as described in Section~\ref{subsec:tools}. If the verification fails, $\mathit{CH}$ can identify the bad signatures with the binary search-based identification method. The main idea is to iteratively split the batch into two sub-batches and verify each of them separately. 
Besides, the sign-then-encrypt scheme is adopted to guarantee the confidentiality, authenticity and integrity of the messages. 

As shown in Fig.~\ref{fig:pro}, in the second layer, $\mathit{CH}$ uploads both $\overline{\mathit{data}}$ and $\mathit{appr}$ to an RSU, which are relayed to $\mathit{CS}$ afterward. Other information sent from $\mathit{CH}$ and the record $\{rc_j\}$ sent from cluster members are related to the \textbf{cluster key audit strategy}, which is designed to further detect the fake data injection attack. The details of the messages, the proposed strategy and the \textbf{bad head identification protocol} are provided in Sections~\ref{subsec:upload} and \ref{subsec:audit}.

%In SADA, the individual sensory data of each vehicle is never revealed to other vehicles, RSUs, or $\mathit{CS}$. Only $\mathit{CH}$ communicates with the RSU so that the identities and trajectories of every member vehicle are not exposed to RSUs or $\mathit{CS}$. The privacy of $\mathit{CH}$ is preserved as well because there is no need to sign $\overline{\mathit{data}}$ with $\mathit{sk}_{i_\text{CH}}$. 
%In addition, 
In SADA, the \textbf{separation of liability} is achieved. To be specific, in the first layer, $\mathit{CH}$ has the responsibility to aggregate an accountable regional average and identify the bad member vehicles (if any). Thus, in the second layer, both RSUs and $\mathit{CS}$ assume the uploaded $\overline{\mathit{data}}$ and $\mathit{appr}$ have been verified by $\mathit{CH}$. 
If $\mathit{CS}$ further detects any invalid or fake $\mathit{appr}$, it is reasonable to believe that $\mathit{CH}$ is an adversary. 
The separation of liability simplifies system management and facilitates the enforcement of potential punishment. 

%% file: 4_details.tex
\section{Framework Design}
\label{sec:details}
In this section, we provide detailed description of each underlying algorithm and protocol. 

\subsection{Recoverable Masking Protocol}
\label{subsec:masking}
Inspired by the work~\cite{bonawitz2017practical}, we adopt a $(t_{\text{sm}}, n_{\text{v}}-1) $ Shamir's secret sharing scheme in the recoverable masking protocol. 
Differently, the objective is to handle bad vehicles rather than dropped users. The revised design is given in Alg.~\ref{alg:alg_mask}:
$v_i$ first generates a construction parameter, $\beta_i$, and a hash of $\beta_i$, $h_i$.
$n_{\text{v}}-1$ shares are generated for $\beta_i$ rather than the Diffie-Hellman secret (different from~\cite{bonawitz2017practical}). 
Both $c_i$ and $h_i$ are sent to an \textit{executor}, who calculates the sum of all sensory data by summing all $c_i$. 
$v_i$ distributes the shares $f_i(j)$ to the corresponding $v_j$. 
Note that, in CluCol, every vehicle performs as an executor. Parameter distribution and exchange are combined with the approval generation process (Section~\ref{subsec:data_approval}).
\begin{figure}[htb!]
	\vspace{-3mm}
	\begin{algorithm}[H]
		\caption{Recoverable Masking Algorithm}
		\label{alg:alg_mask}
		\textbf{Input}: sensory data $\mathit{data}_i$, $p_{\text{mk}}$, $p_{\text{sm}}$, $N_{\text{v}}$, masks $\{\alpha_{ij} \mid  j\in N_{\text{v}} \wedge j\neq i \}$,  threshold $t_{\text{sm}}$, and a hash function $\Hashmask$
		%\textbf{Output}: 
		\begin{algorithmic}[1]
			\State $v_i$ calculates the reconstruction parameter $\beta_i$ as:
			\begin{equation}
			\label{equ:beta_generate}
			\beta_i = \sum_{j\in N_{\text{v}}: i<j}\alpha_{i,j} - \sum_{j \in N_{\text{v}}: i>j}\alpha_{j,i} \mod p_{\text{mk}};
			\end{equation}
			\State Generates the hash of $\beta_i$ as $h_i = \Hashmask (\beta_i)$;
			\State Picks $\{a^2_k \xleftarrow{\S} \mathbb{GF}(p_{\text{sm}})\mid k\in\{1,2,\dots, t_{\text{sm}}-1\} \}$;
			\State Generates $n_{\text{v}}-1$ shares for $\forall j\in N_{\text{v}} \wedge j\neq i$ as
			\begin{equation}
			f_i(j) = \beta_i+\sum_{k=1}^{t_{\text{sm}}-1}a^2_k {j}^k \mod p_{\text{sm}};
			\end{equation}
			\State Masks $\mathit{data}_i$ to $c_i$ as
			\begin{equation}
			\label{equ:equ_mask_alg_recover}
			c_i = \mathit{data}_i + \beta_i \mod p_{\text{mk}};
			\end{equation}
			%	with the masks by Equation~\eqref{equ:equ_masking}. 
			%		as:
			%		\begin{equation}
			%		\label{equ:equ_mask_alg_recover}
			%		c_i = \mathit{data}_i +\sum_{j\in N_{\text{v}}: i<j}\alpha_{i,j} - \sum_{j \in N_{\text{v}}: i>j}\alpha_{j,i};
			%		\end{equation}
			\State \Return $c_i$, $h_i$ and $\{f_i(j)\mid j\in N_{\text{v}} \wedge j\neq i\}$
		\end{algorithmic}
	\end{algorithm}
	\vspace{-5mm}
\end{figure}

When a bad vehicle $v_{i_{\text{re}}}$ is identified, it is required to exclude the sensory data of $v_{i_{\text{re}}}$ and get a new sum. 
To achieve this goal, $\mathit{CH}$ is expected to reconstruct $\beta_{i_{\text{re}}}$ first as follows: $\mathit{CH}$ picks $t_{\text{sm}}$ vehicles except $v_{i_{\text{re}}}$. $\mathit{CH}$ notifies every vehicle of the index $i_{\text{re}}$ and requires the corresponding shares from the picked vehicles. With the $t_{\text{sm}}$ shares, $\beta_{i_{\text{re}}}$ can be recovered by the Lagrange interpolating formula:
\begin{equation}
\beta^{\prime}_{i_{\text{re}}}=\sum_{i \in L_{\text{sm}} } f_{i_{\text{re}}}(i) \prod_{j \in L_{\text{sm}} \wedge j\neq i } \frac{j}{j-i} \mod p_{\text{sm}}.
\end{equation}

$L_\text{sm}$ represents the index list of the picked $t_{\text{sm}}$ vehicles. $p_{\text{sm}}$ is a large prime. 
The reconstructed $\beta^{\prime}_{i_{\text{re}}}$ is sent to each executor. If $\Hashmask (\beta^{\prime}_{i_{\text{re}}}) = h_{i_\text{re}}$, suppose the masked value of $v_{i_{\text{re}}}$ is $c_{i_{\text{re}}}$, the executor can calculate a new sum $\mathit{sum}_{\text{new}}$ from the old sum $\mathit{sum}_{\text{old}}$ as follows:
\begin{equation}
\mathit{sum}_{\text{new}} = \mathit{sum}_{\text{old}}-c_{i_{\text{re}}}+\beta^{\prime}_{i_{\text{re}}} \mod p_{\text{mk}}
\end{equation} 
where $p_{\text{mk}}$ is an integer s.t. $\mathit{data}_i$ and $\sum_{i\in N_\text{v}} \mathit{data}_i \in \mathbb{Z}_{p_{\text{mk}}}$.

Different from the design in~\cite{bonawitz2017practical}, there is no need to recalculate masks. $\beta^{\prime}_{i_{\text{re}}}$ is only reconstructed once by $\mathit{CH}$. 
It obviously saves the computation cost of member vehicles. More comparisons can be found in Section~\ref{subsec:comp}.

\subsection{Data Approval Generation}
\label{subsec:data_approval}

To resist against the fake data injection attack, we introduce a new idea, \textit{data approval}. The principle is to intelligently make minimal modifications to MuSig and satisfy the requirements for the approval generation process.

As a preparation, each vehicle calculates $a^3_i = \Hashagg (L_{\text{pk}}\|\mathit{pk}_i)$ for every vehicle $\{v_i \mid i\in N_{\text{v}}\}$ where $\Hashagg$ is a hash function. The aggregated public key is denoted as $\widetilde{\mathit{pk}}=\prod_{i=1}^{n_{\text{v}}}\mathit{pk}_i^{a^3_i}$.
Each $v_i$ also selects $k_{i} \xleftarrow{\S} \mathbb{Z}_q^{\times}$ and calculates the nonce $R_{i}=g^{k_{i}}$. In MuSig, a signer (i.e., $v_i$) makes a commitment only on $R_{i}$. Differently, we require $v_i$ to generate the commitment $\mathit{com}_i$ as
\begin{equation}
\mathit{com}_{i} = \Hashcom (R_{i}\|c_i \| h_i \| \{ \E_{\mathit{key}_{\text{v}_{i,j}}}(f_i(j))\mid j\in N_{\text{v}} \wedge j\neq i \})
\end{equation}
where $c_i$ is the sensory data masked by Alg.~\ref{alg:alg_mask}. $\Hashcom$ is a hash function. $\E$ denotes an encryption function. Each share $f_i(j)$ is encrypted with a symmetric key of $(v_i, v_j)$. 
Instead of broadcasting $\mathit{com}_{i}$ to all other vehicles in the cluster~\cite{maxwell2019simple}, $v_i$ sends it to $\mathit{CH}$, who generates a 
list of the commitments as $L_{\text{com}}=\{\mathit{com}_{i}\mid i\in N_{\text{v}}\}$.

When and only when receiving $L_{\text{com}}$ from $\mathit{CH}$, $v_i$ sends a message $m_i$ to the whole cluster as:
\begin{equation}
m_i := \{R_{i}\|c_i \| h_i \| \{ \E_{\mathit{key}_{\text{v}_{i,j}}}(f_i(j))\mid j\in N_{\text{v}} \wedge j\neq i \} \}.
\end{equation} 

After receiving $\{m_j\mid j \in N_{\text{v}} \wedge j\neq i\} $ from all other vehicles, $v_i$ first verifies $\Hashcom(m_j) \stackrel{?}{=} \mathit{com}_{j}$ for all $j$. The verification checks the correctness of $m_j$ and prevents an adversary from constructing a fake nonce~\cite{maxwell2019simple}. 
$v_i$ stores both $\E_{\mathit{key}_{\text{v}_{j,i}}}(f_j(i))$ and $h_j$ for potential usage in the recoverable masking protocol and discards other ciphertexts in every $m_j$:  $\{\E_{\mathit{key}_{\text{v}_{j,j^\prime}}}(f_j(j^\prime))\mid j^\prime \in N_{\text{v}} \wedge j^\prime \neq j, i \}$. An aggregated nonce is generated as: $\widetilde{R}=\prod_{j=1}^{n_{\text{v}}}R_j$.

Instead of directly providing each vehicle the average value that requires a Schnorr approval, we ask each $v_i$ to calculate it by itself as:
\begin{equation}
\overline{\mathit{data}} = \frac{1}{n_{\text{v}}}\sum_{j\in N_{\text{v}}}c_j = \frac{1}{n_{\text{v}}}\sum_{j\in N_{\text{v}}} \mathit{data}_i.
\end{equation}
It not only guarantees that no sensory data is exposed to other vehicles, but also confirms the authenticity of the masked data used.
Each $v_i$ now can generate a sub-approval on $\overline{\mathit{data}}$ as $\mathit{appr}_i = (s_{i}, \widetilde{R})$, where $s_{i}=k_{i} + a^3_i \mathit{sk}_i e  \mod q$ and $e = \Hashapp (\widetilde{\mathit{pk}}\|\widetilde{R}\|\overline{\mathit{data}})$.

After collecting all the sub-approvals from member vehicles, $\mathit{CH}$ generates the Schnorr approval for the cluster as $\mathit{appr}=(\widetilde{s}=\sum_{i\in N_{\text{v}}} s_{i} \mod q,\widetilde{R})$.
$\mathit{appr}$ is generated with the contribution of all vehicles so that we can use it as an evidence of the data authenticity.

\subsection{Data Approval Pre-Checking}
\label{subsec:appr_check}
To defend against the input validation attack, $\mathit{CH}$ verifies $\mathit{appr}$ by checking $g^{\widetilde{s}} \stackrel{?}{=} \widetilde{R} \widetilde{\mathit{pk}}^{e^\prime}$ where $e^\prime=\Hashapp(\widetilde{\mathit{pk}}\|\widetilde{R}\| \overline{\mathit{data}})$. We name it as the \textit{first stage}.
If $\mathit{appr}$ is valid, $\mathit{CH}$ can send it to $\mathit{CS}$ on behalf of the whole cluster along with $\overline{\mathit{data}}$. If $\mathit{appr}$ is invalid, $\mathit{CH}$ has the responsibility to identify the invalid sub-approvals in the \textit{second stage}, which can be achieved by verifying all $\mathit{appr_i}$ one by one. A shortcoming of this method is the considerable identification cost.
Another choice is the binary search-based method: $\mathit{CH}$ splits the index list $\{i\mid i\in N_{\text{v}}\}$ to two sublists. For each sublist, $L_\text{sub}$, $\mathit{CH}$ calculates $\widetilde{\mathit{pk}}_{L_{\text{sub}}}= 
\prod_{i \in L_{\text{sub}}}\mathit{pk}_i^{a^3_i} $,  $\widetilde{R}_{L_{\text{sub}}}=\sum_{i \in L_{\text{sub}}}R_i$ and $\widetilde{s}_{L_{\text{sub}}}=\sum_{i \in L_{\text{sub}}}s_i$, and checks $g^{\widetilde{s}_{L_{\text{nd}}}} \stackrel{?}{=} \widetilde{R}_{L_{\text{nd}}} {\widetilde{\mathit{pk}}_{L_{\text{nd}}}}^{e^\prime}$. $\mathit{CH}$ repeats the procedure until all the invalid sub-approvals are found. 

However, the computation cost is still relatively high. Considering that $\mathit{CH}$ cannot bypass the calculation of $\widetilde{\mathit{pk}}$, $\widetilde{s}$ and $\widetilde{R}$ in the first stage, we can maintain the intermediate results at the same time so that no additional calculation is required in both stages. To achieve this goal, we propose the approval pre-checking algorithm with tree structures. $\mathit{CH}$ is expected to construct three binary trees while calculating $\widetilde{\mathit{pk}}$, $\widetilde{s}$ and $\widetilde{R}$: the \textit{public key tree}, the \textit{signature tree} and the \textit{nonce tree}. 

\begin{figure}[htb!]
	\centering
	\vspace{-3mm}
	\includegraphics[width=1.0\linewidth]{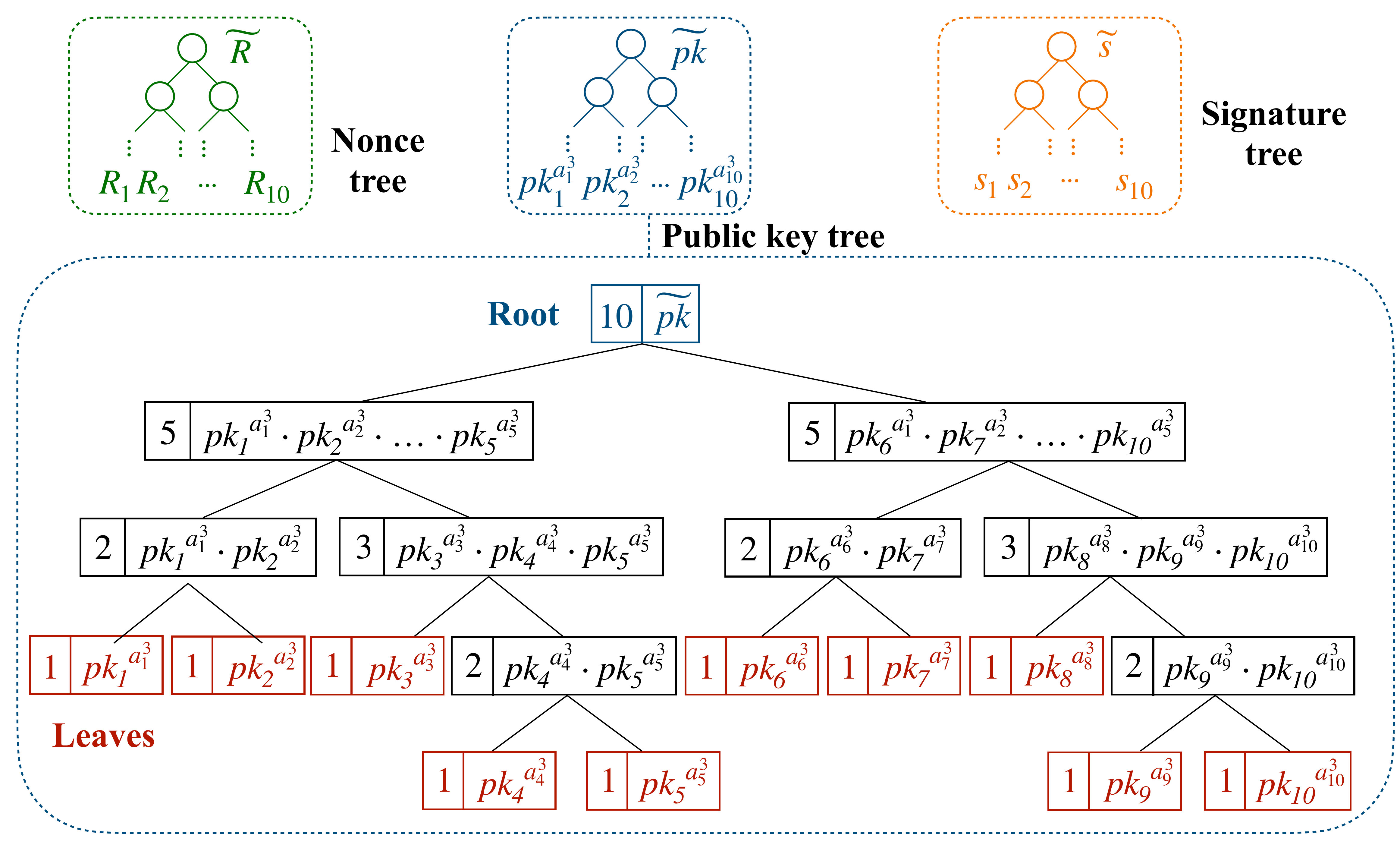}
	\caption{An example of the maintained trees with $n_{\text{v}}=10$.} 
	\label{fig:tree}
\end{figure}

Taking the public key tree as an example, the values of the root node, the $i$-th leaf, and a parent node are $\widetilde{\mathit{pk}}$, $\mathit{pk}_i^{a^3_i}$, and the product of its two children, respectively.
Suppose $L_{\text{nd}}$ is the index list of all the leaves in the subtree of an internal node, we can calculate the value of the node as:
\begin{equation}
\label{equ:pubkey_tree}
\widetilde{\mathit{pk}}_{L_{\text{nd}}} = \widetilde{\mathit{pk}}_\text{left} \cdot  \widetilde{\mathit{pk}}_\text{right} = 
\prod_{i \in L_{\text{nd}}}\mathit{pk}_i^{a^3_i} 
\end{equation}
where $\widetilde{\mathit{pk}}_\text{left} $ and $\widetilde{\mathit{pk}}_\text{right}$ are the values of the left and right children, respectively.
With Alg.~\ref{alg:alg_tree}, $\mathit{CH}$ can construct the public key tree and obtain $\widetilde{\mathit{pk}}$ at the same time.
Similarly, we can calculate the value of an internal node in the signature tree and nonce tree as  $\widetilde{s}_{L_{\text{nd}}}=\sum_{i \in L_{\text{nd}}}s_i$ and $\widetilde{R}_{L_{\text{nd}}}=\prod_{i \in L_{\text{nd}}}R_i$, respectively. An example where $n_{v}=10$ is shown in Fig.~\ref{fig:tree}. 

\begin{figure}[htb!]
	\vspace{-5mm}
	\begin{algorithm}[H]
		\caption{Public Key Tree Generation Algorithm}
		\label{alg:alg_tree}
		\textbf{Input}: $n_{\text{v}}$, $\{\mathit{pk}_i\mid i \in N_{\text{v}}\}$ and  $\{ a^3_i\mid i \in N_{\text{v}}\}$
		%\textbf{Output}: 
		\begin{algorithmic}[1]
			\Procedure {Children}{a parent node $\mathit{node}=(n_{\text{lf}}, 0)$}
			\State \multiline{Generates two children nodes: the left child $\mathit{node}_{\text{l}}=( \lfloor n_{\text{lf}}/2\rfloor, 0)$ and the right child $\mathit{node}_{\text{r}}= (\lceil n_{\text{lf}}/2 \rceil,0)$;}
			\IIf {$\lfloor n_{\text{lf}}/2\rfloor != 1$}
			\textsc{Children} ($\mathit{node}_{\text{l}}$); \EndIIf
			\IIf {$\lceil n_{\text{lf}}/2 \rceil != 1$}
			\textsc{Children} ($\mathit{node}_{\text{r}}$);
			\EndIIf
			\EndProcedure
			
			\State Defines the root node $\mathit{root}$ with a pair $(n_{\text{v}}, 0)$ where the first entry is the number of leaves and the second is the value of the node;
			\State Generates the tree structure by \textsc{Children} ($\mathit{root}$); 
			\State For $\forall i \in N_{\text{v}}$, updates the $i$-th leaf as $(1, \mathit{pk}_i^{a^3_i})$;
			%values of all leaves as: $\{(1, \mathit{pk}_i^{a^3_i})\mid i \in N_{\text{v}}\}\}$;
			\State Updates the values of all other nodes by~\eqref{equ:pubkey_tree};
			\State \Return The public key tree, and the value of $\mathit{root}$
		\end{algorithmic}
	\end{algorithm}
	\vspace{-5mm}
\end{figure}

%The algorithm is concise but subtly combined with the first stage of pre-checking.
With the pre-constructed trees, we fully utilized the inevitable process in the first stage so that there is no need to calculate the intermediate aggregated values in the second stage, which saves the computation cost. 

As introduced in Section~\ref{subsec:masking}, to exclude the bad vehicles who provided invalid sub-approvals, $\mathit{CH}$ first notifies the cluster of $L_{\text{bad}}$, the index list of the bad vehicles, and $L_{\text{sm}}$, the index list of the randomly picked $t_{\text{sm}}$ good vehicles: 
\begin{equation}
m^1_{\text{CH}} := \{L_{\text{bad}}\| L_{\text{sm}} \}.
\end{equation}
$\mathit{CH}$ reconstructs $\{\beta^{\prime}_{i_{\text{re}}} \mid i_{\text{re}}\in L_{\text{bad}}\}$ with the $t_{\text{sm}}$ shares from the picked good vehicles and sends them back to all good vehicles.
Then, every good vehicle can calculate the new average\footnote{A new approval should be generated for the new average. To provide further security, it is suggested to use different nonce $R_i$~\cite{maxwell2019simple}. To save the communication cost, multiple $R_i$ can be transmitted within $m_i$ as a preparation in practice.}.

\subsection{Regional Data Uploading}
\label{subsec:upload}
Each vehicle $v_i$ maintains a credential issued by the TA:
\begin{equation}
\begin{split}
\mathit{cre}_{i} = \{\Sign_{\mathit{ska}}( \Hashvid (\Comm_{\mathit{cka}}(\mathit{ID}_i)),tmp^2) \| \\ \Comm_{\mathit{cka}}(\mathit{ID}_i) \| tmp^2 \}
\end{split}
\end{equation}
where $\Comm_{cka}(\mathit{ID}_i)$ is a cryptography commitment on the identity of $v_i$ with the commitment key of TA, $cka$. 
$tmp^2$ is the expiration time of the credential. $\Hashvid$ is a hash function. The hash is signed with the private key of TA, $ska$.
%The credential is signed by TA with its private key $ska$. 
The credential is different from a traditional public key certificate. 
It does not reveal the identity or public key of $v_i$ but can be used as a commitment to the identity. 

A cluster head is expected to attach its credential $\mathit{cre}_{i_\text{CH}} $ when uploading reports. Beside $\mathit{cre}_{i_\text{CH}}$, as shown in Figure~\ref{fig:pro}, the approval $\mathit{appr} $, the average value $\overline{\mathit{data}}$, the aggregated key $\widetilde{\mathit{pk}}$, and the event ID $\mathit{UID}$ form the report, which should be sent to $\mathit{CS}$ through the RSU. 

To upload the report securely, $\mathit{CH}$ first sends two session keys, encrypted\footnote{For brevity, we do not distinguish the encryption algorithms represented by $\E$. In practice, we can adopt the RSA algorithm to encrypt the session keys and $\overline{\mathit{data}}$ while the AES algorithm in $m^2_{\text{CH}}$ and $m^3_{\text{CH}}$ with $\mathit{key}_1$. } with $\mathit{pkr}$, to the RSU: $\E_{\mathit{pkr}}(\mathit{key}_1, \mathit{key}_2)$. %The RSA algorithm is adopted.
Then, $\mathit{CH}$ sends the following message to RSU:
%Then, to upload $\overline{\mathit{data}}$, $\mathit{CH}$ sends a message to the RSU as:
\begin{equation}
m^2_{\text{CH}}:=\E_{\mathit{key}_1}( \mathit{UID} \| \mathit{appr} \| \E_{\mathit{pks}}(\overline{\mathit{data}}) \| \widetilde{\mathit{pk}}\| \mathit{cre}_{i_\text{CH}} \| \mathit{tmp}^3)
\end{equation}
where $\mathit{tmp}^3$ is a timestamp to defend against the message replay attack. $\overline{\mathit{data}}$ is encrypted with the public key of $\mathit{CS}$ to provide further security.
With $\mathit{appr}$, $\mathit{CS}$ can verify the accountability of $\overline{\mathit{data}}$: it calculates $e^\prime=\Hashapp(\widetilde{\mathit{pk}}\|\widetilde{R}\| \overline{\mathit{data}})$ and checks $g^{\widetilde{s}} \stackrel{?}{=} \widetilde{R} \widetilde{\mathit{pk}}^{e^\prime}$.

\subsection{Cluster Key Audit and Bad Head Identification}
\label{subsec:audit}
One possible attack from a bad $\mathit{CH}$ is to generate a fake $\mathit{appr}$ by itself without the participation of any member vehicle. The fake  $\mathit{appr}$ can be verified successfully by claiming $\mathit{pk}_{i_\text{CH}}$ as $\widetilde{\mathit{pk}}$. 
This attack can be detected by $\mathit{CS}$ with the proposed cluster key audit strategy and bad head identification protocol.

To achieve this goal, we require $\mathit{CH}$ to collect a list of records from all member vehicles, $L_{\text{rc}}$, and upload it to the RSU.
Suppose the total number of the records is $n_{\text{ps}}$ and $N_{\text{ps}}= \{1,2,\dots, n_\text{ps}\}$. We formally define $L_{\text{rc}}$ as follows:
\begin{equation}
L_{\text{rc}}  =\{\mathit{rc}_j=(\Hashvid(\widetilde{\mathit{pk}}_j),\mathit{UID}_j) \mid j\in N_{\text{ps}}\}.
\end{equation}
$\widetilde{\mathit{pk}}_j$ is the aggregated key generated and used in a past data aggregation event with an ID $\mathit{UID}_j$. 
$\Hashvid$ is a hash function.
%the hash value of an aggregated key $\widetilde{\mathit{pk}}_j$, which is generated and used in a past data aggregation event with an ID $\mathit{UID}_j$. 
Each record $\mathit{rc}_j$ in $L_{\text{rc}}$ can be considered as an audit of cluster key $\widetilde{\mathit{pk}}_j$, corresponding to a past event $\mathit{UID}_j$ and a previous cluster head $\mathit{CH}_j$.
Note that during each sensing cycle, $v_i$ only submits the records which have not been uploaded before (typically one record).
Because $L_{\text{rc}}$ is collected from all the vehicles in the cluster, it is impossible for RSUs or $\mathit{CS}$ to infer the trajectory of $\mathit{CH}$ by linking the events. 

To upload $L_{\text{rc}}$ securely, the message is designed as:
\begin{equation}
m^3_{\text{CH}} := \E_{\mathit{key}_1}(\HMAC_{\mathit{key}_2}(L_{\text{rc}} )\| L_{\text{rc}} )
\end{equation}
where $\HMAC$ is a hash-based message authentication code to provide authenticity and integrity.

In each handling cycle, $\mathit{CS}$ compares the hash values in $L_{\text{rc}}$. 
If any record $\mathit{rc}_j$ is invalid, according to the separation of liability principle, $\mathit{CS}$ can believe that the corresponding $\mathit{CH}_j$ is a bad vehicle. 
Thus, even if a bad head uploads a fake approval with the fake key in the current sensing cycle, it can be detected in the upcoming sensing cycles with the information provided by good vehicles. 
The real identity of a bad head can be revoked from $\mathit{cre}_{i_\text{CH}}$ with the help of TA.

%% file: 5_evaluation.tex
\section{Performance Evaluation}
\label{sec:perfor}
In this section, 
we analyze and evaluate the computation and communication costs of SADA, and compare SADA with the latest related works~\cite{li2023eco, kong2019privacy, moni2021scalable, bonawitz2017practical,raja2020sp}, including our previous work, CRS~\cite{liu2023crs}. 
%The latest related works are adopted for comparison~\cite{li2023eco, kong2019privacy, moni2021scalable, bonawitz2017practical,raja2020sp}, as well as our previous work, CRS~\cite{liu2023crs}. 
Following CRS, we consider the scenario where vehicles travel in the same direction along a straight highway. As a representative, $n_\text{v}$ is set to 20 for a cluster. $t_{\text{sm}}$ is set to 10. All simulations are conducted 100 times to get an average result.
\subsection{Computation Cost}
\label{subsec:comp}
\begin{table}[!t]
	\centering
	\small
	\vspace{1.5mm}
	\caption{Cryptographic Operations, Abbreviations and Execution Time~\cite{MIRACL}}
	\label{tab:operation}
	\begin{threeparttable}
		% 	 \resizebox{\textwidth}{!}{
		\begin{tblr}{colspec={X[3.25,l]X[0.8,l]X[1,l]}}
			\toprule
			Operation & Abbr & Time (ms) \\
			\midrule
			Scalar addition\tnote{$*$} & $T_{\text{Z\_ADD}}$ & 0.0037
			\\
			Scalar multiplication\tnote{$*$} & $T_{\text{Z\_MUL}}$& 0.0046
			\\
			Scalar multiplication on $\mathbb{Z}_{p_{\text{n}}}$\tnote{$\dagger$}  & $T_{\text{pn\_MUL}}$ & 1.5922
			\\
			Scalar exponentiation on $\mathbb{Z}_{p_{\text{n}}}$  & $T_{\text{pn\_EXP}}$ & 18.5283
			\\
			Scalar exponentiation on $\mathbb{Z}_{p_{\text{dh}}}$ & $T_{\text{pdh\_EXP}}$ & 26.9556
			\\
			One-way hash function (SHA-256)  & $T_{\text{SHA}}$ & 0.0088
			\\
			Lagrange interpolation & $T_{\text{LI}}$ & 0.1404
			\\
			Point addition on $\mathbb{G}$& $T_{\text{G\_ADD}}$ & 0.0597
			\\
			Point multiplication on $\mathbb{G}$ & $T_{\text{G\_MUL}}$ &9.8134
			\\
			Multiplication on $\mathbb{G}_\text{EG}$\tnote{$\ddagger$} &$T_{\text{EG\_MUL}}$& 0.1628
				\\
			Exponentiation on $\mathbb{G}_\text{EG}$&$T_{\text{EG\_EXP}}$ & 20.2234
			\\
			Logarithm on $\mathbb{G}_\text{EG}$&$T_{\text{EG\_LOG}}$ & 2813.8075
			\\
			RSA-2048 encryption &  $T_{\text{RSA\_E}}$ & 2.8275 
			\\
			RSA-2048 decryption &  $T_{\text{RSA\_D}}$ & 568.1378 
			\\
			AES-256 decryption&  $T_{\text{AES\_DS}}$ & 0.0040 
			\\
			HMAC-SHA256 & $T_{\text{HMAC}}$ & 0.0919
			\\
			Tree construction & $T_{\text{T\_C}}$ & 0.2917
			\\
			Tree traversal& $T_{\text{T\_T}}$ & 0.0015
			\\
			Bos-Coster's algorithm &  $T_{\text{BC}}$ & 170.4289
			\\
			\bottomrule
		\end{tblr}%}
		\begin{flushleft} 
		\begin{tablenotes}
			\footnotesize
			\item[$*$] \textit{Related to $\mathbb{Z}_{p_{\text{mk}}}$, $\mathbb{GF}(p_{\text{sm}})$, and $\mathbb{Z}_q$.}\\
			\item[$\dagger$] \textit{$\mathbb{Z}_{p_{\text{n}}}$ is the multiplicative group used in the Paillier cryptosystem.}\\
			\item[$\ddagger$] \textit{$\mathbb{G}_\text{EG}$ is a cyclic group with a 2048-bit prime order used for ElGamal commitment. Pollard's lambda method~\cite{MIRACL} is used to find discrete logarithms where the exponent is known to be small, e.g., 32 bits~\cite{li2023eco}.}\\
			%\item[$\dagger\dagger$] We consider the time of encrypting 64 bytes (short) and 512 bytes (long) as representative.
		\end{tablenotes}
	 \end{flushleft}
	\end{threeparttable}
	\vspace{-2.5mm}
\end{table}
The execution time of the cryptographic operations simulated with MIRACL~\cite{MIRACL} is listed in Table~\ref{tab:operation}.
The experimental platform is composed of an Intel Q9550 CPU with 2.83 GHz frequency, and 8GB RAM. 
%The operation system is Windows 10 Pro.
We set $\mathbb{G}$, the group used in the Schnorr signature scheme, with the elliptic curve Secp256k1. A 256-bit order $q$ and 2048-bit $p_{\text{dh}}$ are chosen to provide the 128-bit security, where $p_{\text{dh}}$ is the prime order of a Diffie-Hellman group $\mathbb{Z}_{p_\text{dh}}$.
SHA-256 is adopted as the general one-way hash functions. 

%224 bit ECC: 112 bit security level; RSA 2048: 112 bit security level; ECC 256: 128 bit security level

\begin{table*}[hbt!]
	\centering
	\small
	%\vspace{1.5mm}
	\caption{Comparison on the Computation Cost of Secure Data Aggregation}
	\label{tab:compare_confid}	
	\begin{threeparttable}
		\begin{tblr}{colspec={X[0.7,l]X[0.95,l]X[0.98,l]X[1.1,l]X[0.98,l]X[0.4,c]}, rowspec={Q[m]Q[m]Q[m]Q[m]Q[m]Q[m]Q[m]}}
			\toprule 
			Scheme & Method & Data handling (ms) & Aggregation (ms)\tnote{$\P$} & Recover (ms)\tnote{$\|$}  & Accuracy \\
			\midrule
			\SetCell[r=2]{m}SADA & Recoverable masking (in a cluster)  & $(n_{\text{v}}-1+(n_{\text{v}}-1) t_{\text{sm}}) T_{\text{Z\_ADD}}+ 1T_{\text{SHA}}+(n_{\text{v}}-1)(t_{\text{sm}}-1) T_{\text{Z\_MUL}} \approx 1.57$  & $(n_{\text{v}}-1) T_{\text{Z\_ADD}} \approx 0.07$ & {$\mathit{CH}$: $1T_{\text{LI}}+2 T_{\text{Z\_ADD}}$\\$\approx 0.15$; Member vehicle:$1T_{\text{AES\_DS}}+$\\$2T_{\text{Z\_ADD}}\approx 0.01$ } & 	\SetCell[r=2]{m}\cmark
			\\
			& RSA (RSUs-$\mathit{CS}$) & $1 T_{\text{RSA\_E}} \approx 2.83 $ & $1 T_{\text{RSA\_D}} \approx 568.14 $ & NA (Not applicable)&
			\\
			Eco-CSAS~\cite{li2023eco} & ElGamal commitment & {3$T_{\text{EG\_EXP}} $+$1T_{\text{EG\_MUL}} $\\$\approx 60.83$}& $(n_{\text{v}}-1)T_{\text{EG\_MUL}} +1T_{\text{EG\_LOG}}\approx 2816.90$& Not supported  & \cmark
			\\
			Kong's scheme~\cite{kong2019privacy} &  Modified Paillier cryptosystem & $2 T_{\text{pn\_MUL}} + 2 T_{\text{pn\_EXP}}  \approx 40.24$& $2n_{\text{v}} T_{\text{pn\_MUL}} + 1 T_{\text{pn\_EXP}} \approx 82.22$ & Not supported  & \cmark
			%2(n_{\text{v}}-1)T_{\text{pn2\_MUL}} T_{\text{pn2\_MUL}}+ 1 T_{\text{pn2\_EXP}}  + 2T_{\text{pn2\_MUL}} for decrytion
			\\
			SP-CIDS~\cite{raja2020sp} & Differential privacy & 2 $T_{\text{Z\_ADD}} \approx 0.01$ &  $(n_{\text{v}}-1) T_{\text{Z\_ADD}} \approx 0.07$ & Not supported  & \xmark
			\\
			Original design~\cite{bonawitz2017practical} & Masking with threshold secret sharing & $(n_{\text{v}}-1+(n_{\text{v}}-1) t_{\text{sm}}) T_{\text{Z\_ADD}}+(n_{\text{v}}-1)(t_{\text{sm}}-1) T_{\text{Z\_MUL}} \approx 1.56$ &  $(n_{\text{v}}-1) T_{\text{Z\_ADD}} \approx 0.07$ & $1T_{\text{LI}}+(n_\text{v}-2)T_{\text{pdh\_EXP}}+n_\text{v}T_{\text{Z\_ADD}}+ (n_\text{v}-1)T_{\text{SHA}} \approx 485.58$  & \cmark 
			\\ \bottomrule
		\end{tblr}%}
		\begin{flushleft} 
			\begin{tablenotes}
				\footnotesize
				\item[$\P$]\textit{Operations or decryption required to aggregate the data of vehicles by $\mathit{CH}$, RSUs, or the server.}\\
				\item[$\|$] \textit{Reconstruction process and new sum calculation, taking the scenario that one vehicle drops or should be removed as an example. }
			\end{tablenotes}
		\end{flushleft}
	\end{threeparttable}
\end{table*}

To provide confidentiality on the sensory data, each vehicle masks the individual data in clusters. 
%Corresponding parameters and shares are generated as well. 
As shown in Table~\ref{tab:compare_confid}, all the data handling processes in Alg~\ref{alg:alg_mask} take only 1.57ms. 
$\mathit{CH}$ aggregates the shares within 0.07ms and further encrypts the average with the RSA algorithm in 2.83ms.
Compared with the cryptography-based methods~\cite{kong2019privacy, li2023eco}, SADA is more lightweight for vehicles. The heaviest task, i.e., the RSA decryption, is conducted by $\mathit{CS}$ with cloud computing resources so that should not be a bottleneck. 
Although SP-CIDS~\cite{raja2020sp} has a lower data handling cost than SADA, the accuracy of the aggregated data cannot be guaranteed. 
In SADA, $\mathit{CH}$ takes 0.15ms to exclude the data of the bad vehicle and get a new sum with the proposed recoverable masking protocol. A member vehicle only needs 0.01ms to get a new sum. Compared with the works in~\cite{bonawitz2017practical}, where each vehicle takes 485.58ms, our protocol is more lightweight. 
\begin{figure}[!ht]
	\centering
	%\vspace{-3mm}
	\subfloat[]{\includegraphics[width=0.5\columnwidth]{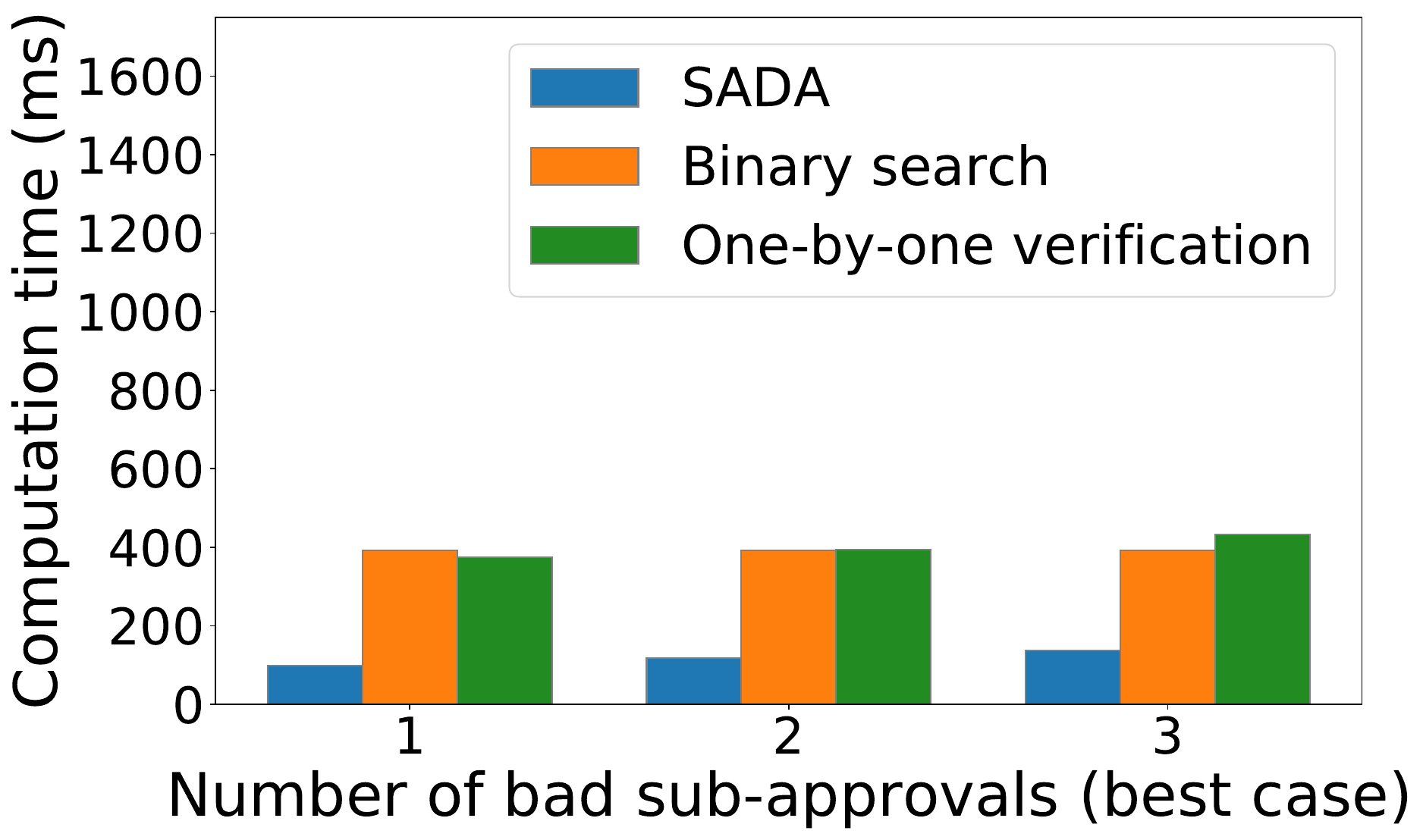}%
		\label{fig:badapproval1}}
	\hfil
	\subfloat[]{\includegraphics[width=0.5\columnwidth]{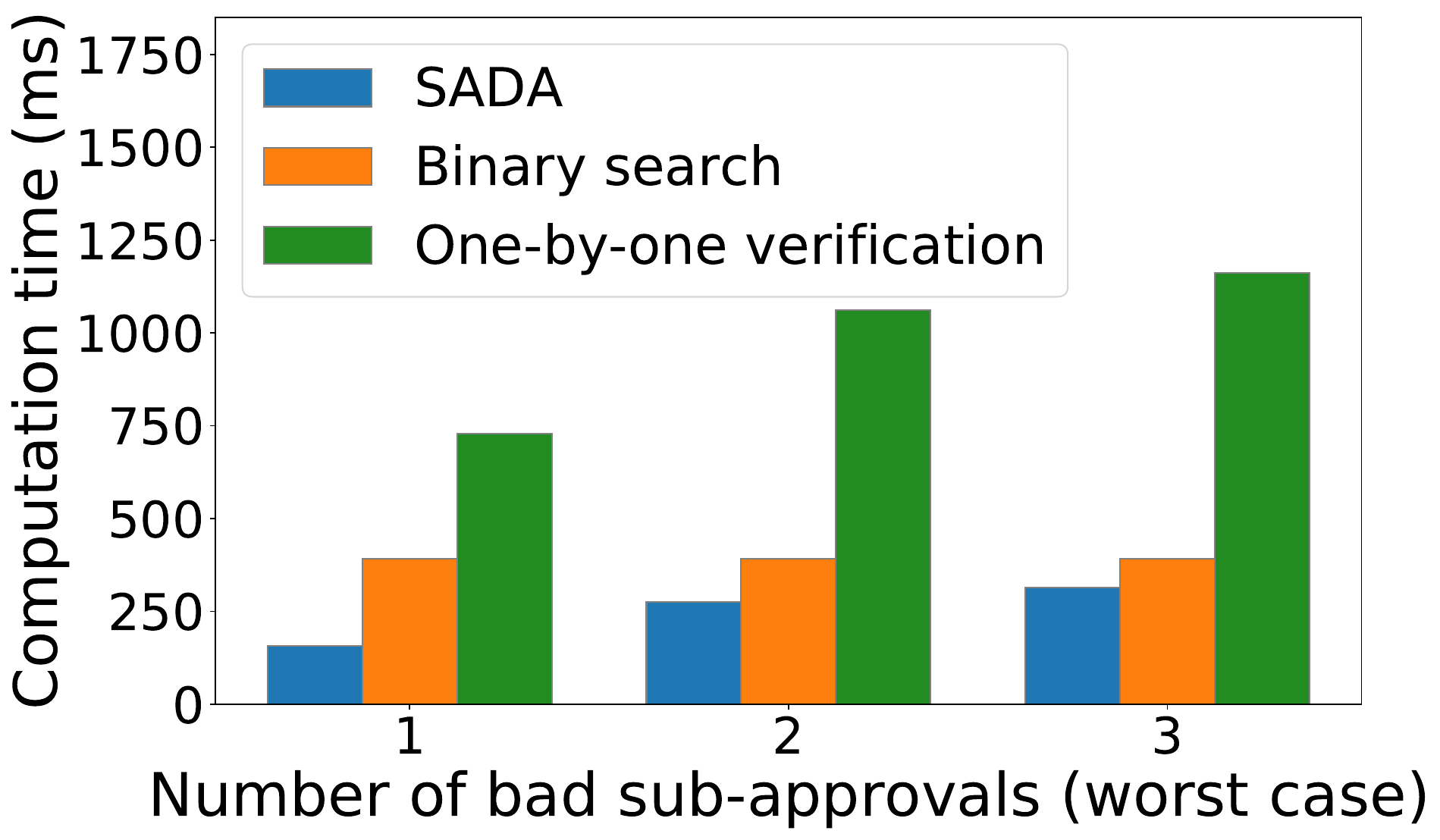}%
		\label{fig:badapproval2}}
	\caption{Comparison of the bad sub-approval identification.}
	\label{fig:badapproval}
	%\vspace{-3mm}
\end{figure}

In data approval pre-checking, we propose an identification method with binary trees. 
We compare it with the naive methods: a) the binary-search method without tree structures, and b) verifying the sub-approvals one by one. 
We set the number of bad vehicles $n_{\text{bad}}=1,2,3$ considering that it should be much less than that of good ones in a cluster.
%It is reasonable that the number of bad vehicles, $n_{\text{bad}}$, is much less than that of good ones. Thus, we set $n_{\text{bad}}$ in a cluster as 1, 2 and 3.
As shown in Fig.~\ref{fig:badapproval}, different from the first naive method, the tree structures save the $O( n_{\text{v}} \log n_{\text{v}} )$  $T_{\text{Z\_ADD}}$ and $T_{\text{G\_ADD}}$ operations required for parameter aggregation. The second method always requires $n_{\text{v}} $ verifications because $\mathit{CH}$ does not have the knowledge of $n_{\text{bad}}$. Overall, the proposed method in SADA works better. 
We briefly introduce the space complexity of the design: there are $2n_\text{v}-1$ nodes in each tree. The space requirement for the three tree structures and the node values are around 9.5KB in total, which is not a bottleneck for vehicles. It is worth noting that only $\mathit{CH}$ is required to maintain the trees.

\begin{table*}[ht!]
	\centering
	\small
	\caption{Comparison on the Cost of Message Verification in VANETs}
	\label{tab:compare_verify}
	\begin{threeparttable}	
		% 	 \resizebox{\textwidth}{!}{
		\begin{tblr}{colspec={X[1,l]X[1.02,l]X[1.4,l]X[1.1,l]X[0.58,l]X[0.55,l]}, rowspec={Q[m]Q[m]Q[m]Q[m]Q[m]}}
			\toprule 
			Scheme & Method & Signing cost (for one message) & Verification cost (for $n_\text{v}-1$ messages)& Signature size (bytes) & Security level (bits)
			\\
			\midrule
			SADA & Schnorr-based batch verification & $1 T_{\text{G\_MUL}} + 1 T_{\text{SHA}} + 1 T_{\text{Z\_MUL}} +1 T_{\text{Z\_ADD}} \approx 9.83$ & $(n_{\text{v}}-1)T_{\text{SHA}} +  1 T_{\text{BC}} \approx 170.60$ & 48  & 128
			\\
			Eco-CSAS~\cite{li2023eco} & Schnorr signature without batch verification & $1 T_{\text{G\_MUL}} + 1 T_{\text{SHA}} + 1 T_{\text{Z\_MUL}} +1 T_{\text{Z\_ADD}} \approx 9.83$& $(n_{\text{v}}-1) (2T_{\text{G\_MUL}} +1 T_{\text{G\_ADD}}+1 T_{\text{SHA}} )\approx 374.21$& 48 & 128
			\\
			Kong's scheme~\cite{kong2019privacy} & Message authentication code & $T_{\text{HMAC}}\approx0.09$ & $(n_{\text{v}}-1)T_{\text{HMAC}}\approx1.75$& 32\tnote{$**$}  &
			256\tnote{$**$} 
			\\
			Moni's scheme~\cite{moni2021scalable}, and CRS~\cite{liu2023crs}  & RSA algorithm & $1 T_{\text{RSA\_D}}\approx 568.14$ & $(n_{\text{v}}-1) T_{\text{RSA\_E}} \approx 53.72$& 256 &112 
			\\ 
			%CRS~\cite{liu2023crs} & RSA algorithm & $1 T_{\text{RSA\_D}}\approx 568.14$ & $(n_{\text{v}}-1) T_{\text{RSA\_E}} \approx 53.72$& 256 &112 
			%\\
			\bottomrule
		\end{tblr}%}
		\begin{flushleft} 
			\begin{tablenotes}
				\footnotesize
				\item[$**$] \textit{32 refers to the size of HMAC rather than a signature. HMAC-SHA256 provides 128-bit collision and 256-bit preimage resistance.}
			\end{tablenotes}	    
		\end{flushleft}
	\end{threeparttable}
\end{table*}

\begin{table*}[ht!]
	\centering
	\small
	\caption{Comparison on the Message Verification and Identity Privacy Preservation (with Our Previous Work, CRS)}
	\label{tab:compare_CRS}
	\begin{tblr}{colspec={X[0.5,l]X[1,l]X[1,l]X[1,l]X[1,l]}, rowspec={Q[m]Q[m]Q[m]Q[m]}}
			\toprule
			Scheme & Verification in cluster & Verification ($\mathit{CH}$-RSU communication) & Privacy of member vehicles & Privacy of $\mathit{CH}$ 
			\\
			\midrule
			SADA & Schnorr-based batch verification & Achieved by Schnorr approvals & \SetCell[r=2]{m}Protected by the two-layered architecture & Achieved by Schnorr approvals
			\\
			CRS~\cite{liu2023crs} & Not supported & RSA algorithm & & Achieved by pseudo-IDs\\
			\bottomrule
	\end{tblr}
\end{table*}

%CRS VS SADA, anonymous signing vs approval 

\begin{table}[ht!]
	\centering
	\small
	\caption{Computation Time of SADA}
	\label{tab:computation}
	\begin{threeparttable}
		% 	 \resizebox{\textwidth}{!}{
		\begin{tabular}{ll}
			\toprule 
			Algorithm or process & Time (ms) \\
			\midrule
			Recoverable masking & 1.57
			\\
			Preparation for approval & 207.38
			\\
			Sub-approval generation  & 55.05
			\\
			Approval generation and tree generation\tnote{$\dagger\dagger$} &  0.97
			%0.8796 (tree)+0.057(appproval aggregation)
			\\
			Approval pre-checking\tnote{$\dagger\dagger$}   & 19.70
			\\
			Invalid sub-approval identification\tnote{$\dagger\dagger$}  & 98.24
			\\
			Reconstruction\tnote{$\dagger\dagger$} &  0.14
			\\
			New sum calculation  & 0.01
			\\
			%Batch verification\tnote{*} & $(n_{\text{v}}-1)T_{\text{SHA}} +  1 T_{\text{BC}} $ & 170.5961
			%\\
			Regional data and records uploading\tnote{$\dagger\dagger$}  & 5.82
			\\ \bottomrule
		\end{tabular}%}
		\begin{tablenotes}
			\footnotesize
			\item[$\dagger\dagger$] \textit{Only performed by $\mathit{CH}$. }
			%\item[**] We mainly focus on the computation cost of vehicles because: 1) $\mathit{CS}$ is deployed on cloud. 2) RSUs usually have more computing resources than vehicles. 3) The burden is a crucial concern of users (i.e., drivers) when they are encouraged to participate a sensing task.
		\end{tablenotes}
	\end{threeparttable}
	\vspace{-3.5mm}
\end{table}

We adopt the Schnorr signatures with batch verification for efficient communication in clusters. As shown in Table~\ref{tab:compare_verify}, it requires low signing time and acceptable verification time.  Although the message authentication code is more efficient, it is not suitable in our scenario because a) it cannot guarantee non-repudiation for messages, and b) additional work is required to distribute a symmetric key between each pair of vehicles. 
CRS does not consider message verification within a cluster but uses RSA signatures with pseudo-IDs for that between cluster heads and RSUs. Compared with SADA, CRS has fewer security guarantees and takes more time in pseudo-ID updating and RSA signing. 
We further compare these two frameworks in Tables~\ref{tab:compare_CRS} and~\ref{tab:security}.

To comprehensively evaluate the computation cost, in Table~\ref{tab:computation}, we further summarize the computation time of all the algorithms and processes in SADA. 
We do not specify explicitly but typically, in a cluster, the communication should be protected with the sign-then-encrypt scheme. Considering that AES-256 is adopted, %Considering that AES-256 is adopted to encrypt the messages transmitted in clusters, 
the total calculation time for a member vehicle and $\mathit{CH}$ is both less than 700ms. If a bad member vehicle exists, to get a new sum and a new approval, around 115ms (a member vehicle) and 515ms ($\mathit{CH}$) are required, additionally and respectively.
The results show that SADA is lightweight for vehicles.

\subsection{Communication Cost}
\label{subsec:comm}
To evaluate the communication cost, we conduct simulations in ns-3.34. The length of a cluster is set to 400m. 
The distance between $\mathit{CH}$ and an RSU is in the range of 0m to 400m. 
We consider $\mathit{CS}$ collects data from a city (e.g., Victoria) or a country (e.g., Canada). Thus the distance between an RSU to $\mathit{CS}$ is 4.48km and 2757km, respectively.
The standard of IEEE 802.11p is adopted for both V2V (vehicle-to-vehicle) and V2I (vehicle-to-infrastructure) communications. The main communication cost is analyzed and simulated in sequence as follows:

In a cluster, each $v_i$ sends $\mathit{com}_i$ to $\mathit{CH}$, which is 32 bytes with SHA-256. The returned list $L_\text{com}$ is considered as $20\times32=640$ bytes. Considering that each share can be represented with 2 bytes and encrypted with AES-256, the length of $m_i$ is around 370 bytes. Overall, there are $2(n_{\text{v}}-1)+n_{\text{v}}(n_{\text{v}}-1)$ messages sent in a cluster as a preparation for approval generation. Sending $\mathit{com}_i$, $L_\text{com}$ and $m_i$ takes 0.33ms, 1.93ms, and 0.79ms, respectively. Each sub-approval (48 bytes) is transmitted in 0.35ms.

When a bad sub-approval is detected, $t_{\text{sm}}$ vehicles send the corresponding shares to $\mathit{CH}$. The reconstructed $\mathit{\beta^{\prime}_{i_{\text{re}}}}$ is 2 bytes. There is no need for vehicles to re-exchange masks where $\mathcal{O}(n_{\text{v}})$ messages should be sent for each vehicle. With the proposed protocol, only $\mathcal{O}(1)$ messages are required between each vehicle and $\mathit{CH}$. Thus, the communication cost is low (around 0.31ms). 
%There is no need for vehicles to re-exchange masks or re-conduct the masking process. Thus, the communication cost is low (around 0.31ms). 

Recall that the communication in a cluster is protected with the Schnorr signature and AES-256. All the above results are simulated with considering the size of signatures and ciphertexts. We further compare the signature size in different VANETs-related works in Table~\ref{tab:compare_verify}. 
The adopted Schnorr signature is obviously smaller than that of the RSA signature scheme.
Overall, the V2V communication in a cluster is lightweight. 

As for the V2I communication, considering $\mathit{CH}$ traveling at 20m/s, it takes 0.56ms on average to transmit the session keys from $\mathit{CH}$ to the RSU. Then, to upload the aggregated data and $L_\text{rc}$, two messages are sent from $\mathit{CH}$: 
$m^2_{\text{CH}}$ (864 bytes) and $m^3_{\text{CH}}$ (1024 bytes).
They cost 2.14ms and 2.37ms, respectively. 
Suppose the same sign-then-encrypt scheme is adopted between RSUs and $\mathit{CS}$, 
the message sent takes 0.13ms and 56.33ms for citywide data aggregation and national data aggregation, respectively. 

It is worth emphasizing that, different from the pseudonym-based schemes~\cite{moni2021scalable,liu2023crs}, SADA does not require vehicles to acquire and update pseudonyms with TA, which further saves the communication cost for vehicles.

In this work, we do not consider cluster changes in a sensing cycle.
We can further narrow the assumption down as: the cluster is stable in the process of regional data aggregation. The required time is around 1.5s (considering the presence of bad member vehicles) with the previous analysis and simulation. In this case, the assumption is reasonable in practice with the many existing techniques. 
In addition, we do not consider message loss in the vehicle network.
We refer readers to our previous work~\cite{liu2023crs} for more related contents where a) we discuss more about the stability of clusters and introduce the state-of-the-art research; b) we conduct simulations with more vehicle mobility settings and real-world datasets, which impact the message loss rate and recovery performance.

%to our previous work~\cite{liu2023crs} for more discussion about cluster stability.

%% file: 6_security.tex
\section{Security Analysis}
\label{sec:sec}
%In \textbf{\textit{Summary}} 
\begin{table*}[ht!]
	\centering
	\small
	%\vspace{-0.8mm}
	\caption{Comparison on Security and Privacy Protection}
	\label{tab:security}
	\begin{threeparttable}
		% 	 \resizebox{\textwidth}{!}{
		\begin{tblr}{colspec={X[1,l]X[0.9,l]X[1.15,l]X[1,l]X[0.8,l]X[1,l]}, rowspec={Q[m]Q[m]Q[m]Q[m]Q[m]Q[m]Q[m]Q[m]}}
			\toprule
			Property & SADA & CRS~\cite{liu2023crs}  & Kong's scheme~\cite{kong2019privacy} &  SP-CIDS~\cite{raja2020sp} & Eco-CSAS~\cite{li2023eco} \\
			\midrule
			IoV data confidentiality & \cmark (Masking) & \cmark (Masking) & \cmark (Modified Paillier) & \cmark (Differential privacy) & \cmark(ElGamal commitment)
			\\
			Identity privacy & \cmark (Cluster and approval) & \cmark (Cluster and pseudonym)& \xmark & \xmark & \cmark(Pseudonym)
			\\
			Trajectory privacy  & \cmark (Cluster and approval) & \cmark (Cluster and pseudonym)
			& \halfcheckmark\tnote{$\dagger\dagger$}  & \xmark & \cmark(Pseudonym)
			\\
			IoV data authenticity & \cmark (Approval)& \xmark & \xmark & \xmark & \xmark
			\\
			Approval validity & \cmark (CluCol) & NA & NA & NA & \cmark (Zero-knowledge proof)\tnote{$\ddagger\ddagger$}
			\\
			Malicious head or RSU detection\tnote{$\S\S$} & \cmark (CluCol) & \xmark & \xmark & \xmark & \cmark (Blockchain)
			\\
			Message authentication and integrity & \cmark (Digital signature and HMAC) & \halfcheckmark (Digital signature only for $\mathit{CH}$-RSU communication) & \cmark (Message authentication code) & \xmark & \cmark (Digital signature)
			\\
			%Batch verification &\cmark (Schnorr) & \xmark & \xmark &\xmark & \xmark\\
			 \bottomrule
		\end{tblr}%}
		\begin{flushleft} 
			\begin{tablenotes}
				\footnotesize
				\item[$\dagger\dagger$] \textit{The location information maintained in the report is protected but the trajectory can be inferred by linking the communications.} \\
				\item[$\ddagger\ddagger$]\textit{The scheme can prove that the ElGamal commitment is well-formed.}\\
				\item[$\S\S$] \textit{In SADA, we assume RSUs firmly follow the protocol but a bad $\mathit{CH}$ may perform a fake data injection attack. It can be prevented and detected by CluCol. In Eco-CSAS, there is no cluster but the authors consider that a malicious RSU may delete or tamper some critical data it collects. It can be detected with a blockchain where all the data are stored.}
			\end{tablenotes}
		\end{flushleft} 
	\end{threeparttable}
	\vspace{-3.5mm}
\end{table*}

In Table~\ref{tab:security}, we further compare SADA with the related works focusing on secure data aggregation in VANETs. It shows that, considering both the security and privacy protection performance and the computation and communication costs, SADA is more suitable for IoV data aggregation.
In the following subsections, we analyze the security protection and privacy preservation of SADA against the typical attacks and potential behaviors of the adversaries.

\subsection{Data Leakage Attack}
The individual sensory data are masked with uniformly random masks.
With~\eqref{equ:beta_generate} and~\eqref{equ:equ_mask_alg_recover}, Lemma~\ref{lemma_mask} shows that the masked values $\{c_i\mid i \in N_{\text{v}}\}$ look uniformly random as well~\cite{bonawitz2017practical,song2022eppda}.
\begin{lemma}
	\label{lemma_mask}
	Fix $N_\textup{v}$, $p_\textup{dh}$,  $p_\textup{mk}$ where $p_\textup{dh}>p_\textup{mk}$, and $\{\mathit{data}_i\mid i\in N_\textup{v}\}$ where $\forall i \in N_\textup{v}$, $\mathit{data}_i \in \mathbb{Z}_{p_\textup{mk}}$. Then,
	\vspace{-1mm}
	\begin{equation}
	\begin{split}
	\{ \{\alpha_{i,j} \xleftarrow{\S} \mathbb{Z}_{p_\textup{dh}} \mid i<j \}, \{\alpha_{j,i} \xleftarrow{\S} \mathbb{Z}_{p_\textup{dh}}\mid i>j \}:\\ \{\mathit{data}_i+\beta_i \mod p_\textup{mk} \mid i\in N_\textup{v} \}\} \\ \equiv \{\{c_i \xleftarrow{\S} \mathbb{Z}_{p_\textup{mk}}\mid i\in N_\textup{v}\}: \{c_i\mid i\in N_\textup{v}\} \}
	\end{split}
	\end{equation}
	where $\equiv$ denotes the identical distribution.
\end{lemma}

With Lemma~\ref{lemma_mask}, the probability that any adversary, who obtains any $c_i$ of interest, infers the corresponding $\mathit{data}_i$ is equal to the probability imposed by its a-priori knowledge.

\subsection{Identity Leakage Attack and Trajectory Tracking Attack}

In SADA, member vehicles are hidden behind $\mathit{CH}$. There is no direct communication between a member vehicle and the semi-honest $\mathit{CS}$. There is no need to share the identities or public keys of member vehicles to $\mathit{CS}$. In addition, for any adversary $\mathcal{A}$, inferring $\mathit{pk}_i$ from $\mathit{\widetilde{\mathit{pk}}}$ is as hard as a random guess~\cite{maxwell2019simple}. 
Thus, the identity and trajectory privacy of member vehicles are preserved. 
%The probability that $\mathit{pk}_i$ for any $i$ and any $j$

%in standard model
The identity privacy of $\mathit{CH}$ relies on the security of the commitment scheme. %$\mathit{cre_{i_\text{CH}}}$ only contains the commitment of $\mathit{ID_{i_\text{CH}}}$.
The Pedersen commitment scheme is perfect hiding in \textbf{standard model} with the \textbf{discrete logarithm (DL) assumption}: 
%given the committed identity $\Comm_{cka}(\mathit{ID}_{i_{\text{CH}}})$, every possible $\mathit{ID}_i$ is equally likely to be the value $\mathit{ID}_{i_{\text{CH}}}$ in the commitment, i.e.,
for $\forall i, i^\prime \in\{1,2,\dots, n_\text{ID}\}$ where $n_\text{ID}$ is the number of registered IDs in the system, $P(\mathit{ID}_i=\mathit{ID}_{i_{\text{CH}}} \mid \Comm_{cka}(\mathit{ID}_{i_{\text{CH}}})) = P(\mathit{ID_i^\prime}=\mathit{ID}_{i_{\text{CH}}} \mid \Comm_{cka}(\mathit{ID}_{i_{\text{CH}}})) $. Even for an all-powerful adversary, without the commitment key of TA, $\mathit{cka}$, the probability that it can infer $i_{\text{CH}}$ from $\Comm_{cka}(\mathit{ID}_{i_{\text{CH}}})$ is negligible larger than a blind guess. Besides, guessing $\mathit{pk}_{i_{\text{CH}}}$ from the aggregated key is as hard as a random guess, which is meaningless. Thus, both RSUs and $\mathit{CS}$ cannot infer the real identity or public key of $\mathit{CH}$. 

The protection of the trajectory of $\mathit{CH}$ is twofold: 1) the roles of vehicles are changing between sensing cycles. The probability that $\mathit{CH}$ is elected as the cluster head in next $n_\text{c}$ continuous sensing cycles, is as low as $\frac{1}{{n_\text{v}}^{n_\text{c}}}$. It is hard for RSUs to trace the vehicle by linking the credential. 2) In practice, $\mathit{CH}$ updates $\mathit{cre}_{i_\text{CH}}$ with TA after each use or a pre-determined time threshold, depending on the privacy requirement.

\subsection{Fake Data Injection Attack}
\label{subsec_security_fdia}
To achieve the fake data injection attack, the intuitive way for the adversary, i.e., the bad $\mathit{CH}$, is to provide a fake $\overline{\mathit{data}}$ to the whole cluster and ask for a valid approval. This is defended against by the design: $\overline{\mathit{data}}$ is calculated by every vehicle rather than directly provided by $\mathit{CH}$.

Another potential method is \textit{to forge a cluster approval on behalf of the whole cluster}. This attack is defended against because the approval scheme is provably secure under the \textbf{DL assumption} in the \textbf{plain public-key model}.

\begin{theorem}
	\label{theorem_forg}
	Assume there exists a $\mathit{CH}$ which is a $(T_{\textup{fg}}, n_{\textup{qs}}, n_{\textup{qh}}, n_{\textup{v}}, \epsilon)$ forger against the approval generation algorithm with $p_\text{mk}$ and group parameters $(\mathbb{G}, q, g)$, where $q$ is $l_\textup{q}$-bit long. Assume the hash functions $\Hashcom, \Hashagg, \Hashapp: \{0,1\}^*\rightarrow \{0,1\}^{l_\textup{hash}}$ are modeled as random oracles. 
	Then, there exists an algorithm $\mathcal{C}$ which $(T_{\mathcal{C}}, \mathit{acc}(\mathcal{C}))$-solves the DL problem for $(\mathbb{G}, q, g)$, with $	T_{\mathcal{C}} = 4T_\textup{fg} + 4\mathcal{O}(n_\textup{v}(n_{\textup{qs}}+n_{\textup{qh}})) + 4n_\textup{v}T_\textup{G\_MUL} $ and $\mathit{acc}(\mathcal{C}) \geq \frac{\epsilon^4}{(n_{\textup{qs}}+n_{\textup{qh}}+1)^2(n_{\textup{qs}}+n_{\textup{qh}})} -8(\frac{p_\textup{mk}+1}{p_\textup{mk}})\frac{n_{\textup{qs}}(n_\textup{v}n_{\textup{qs}}+n_{\textup{qh}})}{2^{l_\textup{q}}} - \frac{16(n_\textup{v} n_{\textup{qs}} + n_{\textup{qh}})^2+3}{2^{l_\textup{hash}}}$. 
\end{theorem}

Theorem~\ref{theorem_forg} indicates that if a bad $\mathit{CH}$ in ROM runs in time at most $T_{\text{fg}}$, initiates at most $n_{\text{qs}}$ signature protocols and at most $n_{\text{qh}}$ hash queries to each of the random oracles, and forges a valid cluster approval on behalf of a cluster ($n_\text{v}$ vehicles) with probability at least $\epsilon$, then the corresponding DL problem can be solved in time $T_{\mathcal{C}}$ with probability at least $acc(\mathcal{C})$.  

To prove Theorem~\ref{theorem_forg}, we can follow the same strategy and use the double forking lemma as the original proof of MuSig~\cite{maxwell2019simple}. However, because the design of the approval scheme is different from that of MuSig, there are obvious changes in the detailed proof process and results. 
We provide the details in Appendix. It shows that the adjustments and revision on MuSig in the approval generation algorithm do not affect the security property but result in different $T_{\mathcal{C}}$ and $acc(\mathcal{C})$.

Now we discuss another method of the adversary to implement the attack: $\mathit{CH}$ may \textit{generate a fake but valid approval with its own public key $\mathit{pk}_{i_{\textup{CH}}}$ and claim the key as the cluster key}. 
We define the security game as follows: 
\begin{game}
	In a data aggregation event $\mathit{UID}_j$, the adversary generates an approval only with its own public key $\mathit{pk}_{i_\textup{CH}}$.
	The adversary uploads the approval, following the protocol of SADA, but provides $\mathit{pk}_{i_\textup{CH}}$ to $\mathit{CS}$ instead of the real cluster key $\widetilde{\mathit{pk}}_j$.
	The adversary wins if $\mathit{CS}$ cannot detect the fake approval with the cluster key audit strategy, i.e., $\textup{Count}(\mathit{EventA}) < t_\textup{aud}$. 
	The event $\mathit{EventA}$ is true when $\mathit{CS}$ receives a record corresponding to $\mathit{UID}_j$ showing that $\Hashvid(\mathit{pk_{i_\text{CH}}})\neq \Hashvid(\widetilde{\mathit{pk}}_j)$. The $\textup{Count}$ function counts the number of times that $\mathit{EventA}=\mathit{true}$. $t_\textup{aud}$ is a pre-defined system threshold which satisfies $t_\textup{aud}\geq 1$. 
\end{game}

The adversary wins the game when and only when $\mathit{pk_{i_\text{CH}}}$ occasionally leads to the same hash value of $\widetilde{\mathit{pk}}_j$.
Thus, the security relies on the underlying hash function. We adopt SHA-256 which provides a 128-bit security level against collision and preimage attacks so that the probability that the adversary wins the game is negligible.

\subsection{Input Invalidation Attack}
The input validation attack can be detected with the proposed data approval pre-checking algorithm. 
It is infeasible for an adversary to find an invalid sub-approval $(s_i^\prime, \widetilde{R})$ that satisfies $s_i^\prime \neq s_i$ but $g^{(\sum_{j\in N_v \wedge j\neq i}s_j)+s_i^\prime} = \widetilde{R} \widetilde{\mathit{pk}}^{e^\prime}$. 
The proof is straightforward: $g^{(\sum_{j\in N_v \wedge j\neq i}s_j)+s_i^\prime} = g^{(\sum_{j\in N_v \wedge j\neq i}s_j)+s_i}$ indicates that $s_i=s_i^\prime$.

%% file: 7_conclu.tex
\section{Conclusion}
\label{sec:con}
In this paper, we proposed a Schnorr approval-based IoV data aggregation framework.
%we propose a regional data collection scheme for vehicular clusters. 
A cluster head can aggregate the sensory data in a privacy-preserving way.
Invalid sub-approvals can be identified efficiently and a new average can be easily calculated.
 %The cluster head can verify the generated approval and identify the invalid sub-approvals efficiently. A new average can be easily calculated when a bad vehicle is identified. 
%The scheme is further extended to a secure data aggregation framework. 
With the approval, the authenticity of the aggregated data can be verified and there is no need for pseudonyms, which addresses the security and efficiency limitations of CRS~\cite{liu2023crs}. Both the identities and trajectories of vehicles are protected. Compared with the related works, our work not only meets more security requirements but also is lightweight for vehicles.

%% file: 8_appendix.tex
\appendix[Proof of Theorem~\ref{theorem_forg}]
%\label{app:proof}
\renewcommand\theequation{A\arabic{equation}}
We follow the same strategy as the original proof of MuSig~\cite{maxwell2019simple}.
For brevity, we omit the same definitions and some details of the proof but highlight the difference from that of MuSig. 

\renewcommand\qedsymbol{$\blacksquare$}

\begin{proof}
	We first construct an algorithm $\mathds{A}$ simulating the oracles, running the forger ($\mathit{CH}$), answering the queries, and returning a forgery unless some bad events happen. 
	We assume there is a honest member vehicle $v_{i^*}$ with the key pair $(\mathit{sk}_{i^*}, \mathit{pk}_{i^*})$. The input of $\mathds{A}$ includes $\mathit{pk}_{i^*}$, and uniformly random strings $h_{0,1}, h_{0,2}, \dots, h_{0, n_\text{q}-1}$ and $h_{1,1}, h_{1,2}, \dots, h_{1, n_\text{q}}$ where $n_\text{q}=n_{\text{qs}}+ n_{\text{qh}}+1$.

	\begin{queryalg}[ht!]
		\caption{Approval Query $(L_\text{pk}, c_{i^*}, \mathit{info}_{i^*}) $}
		\label{alg:app_q}
		%\textbf{Input}: index of the vehicle who is expected to be excluded $i_{\text{re}}$, large prime number $p_{\text{sm}}$, threshold $t_{\text{sm}}$
		%\textbf{Output}: 
		\begin{algorithmic}[1]
			\IIf {$\mathit{pk}_{i^*}  \centernot\in L_\text{pk}$} 
				\Return $\perp$.
			\EndIIf
			\State Supposes $L_\text{pk} = \{\mathit{pk}_{i^*}, \mathit{pk}_2,\dots, \mathit{pk}_{n_\text{v}}\}$, i.e., $i^*=1$;
			\If{$T_\text{agg}(L_\text{pk}\| \mathit{pk}_1)$ is undefined}
			\State Internal query: hash query $\Hashagg(L_\text{pk}\| \mathit{pk}_1)$;
			\EndIf
			\State Sets $a^3_i = T_\text{agg}(L_\text{pk}\| \mathit{pk}_{i})$ for each $i$; $\widetilde{\mathit{pk}} = \prod_{i=1}^{n_\text{v}} \mathit{pk}_i^{a_i^3}$;
			\State Sets $\mathit{ctr}_1++$ and sets $e = h_{1,\mathit{ctr}_1}$; 
			\State Generates a random value as the signature $s_1 \xleftarrow{\$} \mathbb{Z}_q$;
			\State Calculates $R_1$ from $s_1$: $R_1=g^{s_1(\mathit{pk}_1)^{-{a^3_1}e}}$;
			\If{ $T_\text{com}(R_1\| c_1\| \mathit{info}_1)$ is defined }
			\State Sets the flag $\mathit{BadCom}_1$ to true;
			\Return $\perp$.
			\EndIf
			\State Internal query: hash query $\Hashcom(R_1\| c_1\| \mathit{info}_1)$;
			\State Sets $\mathit{com}_1= T_\text{com}(R_1\| c_1\| \mathit{info}_1)$ and sends it to $\mathit{CH}$;
			\State Receives $L_{\text{com}}=\{\mathit{com}_{i}\mid i\in N_{\text{v}}\}$ collected by $\mathit{CH}$;
			\State For $\forall i \in N_{\text{v}} \wedge i\neq 1$, looks for entries $(R_i\| c_i \| \mathit{info}_i)$ s.t. $T_\text{com}(R_i\| c_i\| \mathit{info}_i)= \mathit{com}_i$;
			
			\If {For some $i$, more than one entry can be found}
			%i.e., two $T_\text{com}(R_i\| c_i\| \mathit{info}_i)$ have the same value}
			\State Sets the flag $\mathit{BadCom}_2$ to true;
			\Return $\perp$.
			\EndIf
			
			\If {For some $i$, no such value can be found}
			%, i.e., there is no entry for some $\mathit{com}_i$ in $T_\text{com}$}
			\State Sends $m_1 = \{R_1\| c_1\| \mathit{info}_1\}$ to $\mathit{CH}$;
			\State Receives $m_i = \{R_i\| c_i\| \mathit{info}_i\}$ for $\forall i\in N_{\text{v}} \wedge i\neq 1$;
			\State Internal query: hash query $\Hashcom(R_i\| c_i\| \mathit{info}_i)$ for these $i$;
			\If {For some $i$, $\Hashcom(R_i\| c_i\| \mathit{info}_i) \neq \mathit{com}_i$ }
			\State Aborts the protocol.
			\Else
			%\ElsIf{For all $i$, $\Hashcom(R_i\| c_i\| \mathit{info}_i) = \mathit{com}_i$}
			%, i.e., the randomlly assigned value for the missed entry is the same with $\mathit{com}_i$} 
			\State Sets the flag $\mathit{BadCom}_3$ to true;
			\Return $\perp$.
			\EndIf
			\EndIf
			
			\If {Exactly one entry can be found for each $i$}
			\State Sends $m_1 = \{R_1\| c_1\| \mathit{info}_1\}$ to $\mathit{CH}$;
			\State Receives $m_i = \{R_i\| c_i\| \mathit{info}_i\}$ for $\forall i\in N_{\text{v}} \wedge i\neq 1$;
			%\State Checks if $\Hashcom(R_i\| c_i\| \mathit{info}_i) = \mathit{com}_i$
			%\If {If not all $\Hashcom(R_i\| c_i\| \mathit{info}_i) = \mathit{com}_i$} 
			\If {For some $i$, $\Hashcom(R_i\| c_i\| \mathit{info}_i) \neq \mathit{com}_i$ }
			\State Aborts the protocol.
			\EndIf
			\State Computes $\widetilde{R} =\prod_{i=1}^{n_\text{v}} R_i$;
			\State Computes the average value $\overline{data} = \frac{1}{n_\text{v}} \sum_{i=1}^{n_\text{v}} c_i$;
			\If {$T_\text{app}(\widetilde{\mathit{pk}}\| \widetilde{R}\| \overline{data})$ has already been defined}
			\State Sets $\mathit{BadProg}$ to true;
			\Return $\perp$.
			\Else
			%\ElsIf{$T_\text{app}(\widetilde{\mathit{pk}}\| \widetilde{R}\| \overline{data})$ is not defined}
			\State Assigns $T_\text{app}(\widetilde{\mathit{pk}}\| 	\widetilde{R}\| \overline{data})= e$;
			\EndIf
			\State Sends $s_1$ to the forger, i.e., $\mathit{CH}$.	
			\EndIf
		\end{algorithmic}
	\end{queryalg}
	
	Algorithm $\mathds{A}$ is responsible for answering four kind of queries to $\mathit{CH}$: 1) hash query $\Hashcom(R_i\| c_i\| \mathit{info}_i)$ where $\mathit{info}_i$ denotes $h_i\| \{\E_{\mathit{key}_{\text{v}_{i,j}}}(f_i(j))\mid j\in N_\text{v} \wedge j\neq i \}$. $\mathds{A}$ randomly assigns a value to the corresponding hash table entry, $T_\text{com}(R_i\| c_i\| \mathit{info}_i) \Randassignl$, if it is undefined, and returns $T_\text{com}(R_i\| c_i\| \mathit{info}_i)$. 
	2) Hash query $\Hashagg(L_\text{pk}\| \mathit{pk}_i)$. 3) Hash query $\Hashapp(\widetilde{\mathit{pk}}\| \widetilde{R} \| \overline{data} )$. These two queries remain the same as that in MuSig~\cite{maxwell2019simple}. Corresponding hash tables are assigned randomly or directly from the input strings. Corresponding counters, $\mathit{ctr}_0$ and $\mathit{ctr}_1$, are increased. The queried hash values are returned as well. 4) Approval query, which simulates the sub-approval generation oracle as the honest cluster member $v_{i^*}$. We describe it in Query~\ref{alg:app_q} and highlight the difference from the signature query in MuSig as follows:
	\begin{itemize}
		\item In MuSig, the message that requires signing is given explicitly as input. However, in SADA, the average value is calculated by every vehicle in the cluster independently. 
		\item We must exchange $m_i$ first, calculate $\overline{data}$ (Steps 31--37), and then check whether $T_\text{app}(\widetilde{\mathit{pk}}\| \widetilde{R}\| \overline{data})$ is already defined (Steps 38--42). 
		\item Because of the different logic in Query~\ref{alg:app_q}, there is no need to remain the flag $\mathit{Alert}$ which is used along with the flag $\mathit{BadCom}_3$ in MuSig.
	\end{itemize} 
	
	If the forger returns a forgery $(\widetilde{R},\widetilde{s})$, $\mathds{A}$ checks the validity of the forgery as described in the proof of MuSig~\cite{maxwell2019simple}. Two bad events, $\mathit{BadOrder}$ and $\mathit{KeyColl}$, may occur during this process. The major differences are as follows:
	\begin{itemize}
		\item Besides the corresponding $L_\text{pk}$ of the forgery $(\widetilde{R},\widetilde{s})$, the aggregated key $\widetilde{\mathit{pk}}$ is returned by the forger as well. We assume that the returned $\widetilde{\mathit{pk}}$ corresponds the set $L_\text{pk}$ where $\mathit{pk}_{i^*}\in L_\text{pk}$. This assumption is reasonable because we are analyzing the situation where the bad $\mathit{CH}$ tries to generate a forgery on behalf of the whole cluster. In other words, it tries to upload a fake approval with the real aggregated public key.
		\item To return $\widetilde{\mathit{pk}}$, the forger must make a direct or internal $\Hashagg$ query. Besides, there is no need for  $\mathds{A}$ to calculate $\widetilde{\mathit{pk}}$ and check it. Thus, there is no need to check whether $T_\text{agg}(L_\mathit{pk}\| \mathit{pk}_{i^*})$ is defined. 
		This modification leads to the different number of $\Hashagg$ queries in $\mathds{A}$. 
	\end{itemize} 
	
	Now we prove the lower bound of the accepting probability of $\mathds{A}$.
	Suppose $\mathit{CH}$ is a $(T_{\text{fg}}, n_{\text{qs}}, n_{\text{qh}}, n_{\text{v}}, \epsilon)$ forger, then the probability that $\mathds{A}$ returns a forgery is as follows:
	\begin{equation}
		\label{eq_prob_A_1}
		\mathit{acc}(\mathds{A}) \geq \epsilon-\Prb[\text{Bad}]
	\end{equation}
	where $\Prb[\text{Bad}]$ is the probability that the bad events happen. $\mathit{BadCom}_1$ happens when $T_\text{com}(R_1\| c_1\| \mathit{info}_1)$ has been defined. $\mathit{BadCom}_2$ happens when two entries in $T_\text{com}$ have the same value. $\mathit{BadCom}_3$ happens when some $\Hashcom(R_i\| c_i\| \mathit{info}_i)$ is undefined and gets set by chance to the value $\mathit{com}_i$. 
	The probabilities remain the same as that in~\cite{maxwell2019simple}:
	\begin{equation}
		\label{eq_prob_A_2}
		\Prb[\mathit{BadCom}_1] \leq n_{\text{qs}}(n_\text{v}n_{\text{qs}}+ n_{\text{qh}})/2^{l_\text{q}-1},
	\end{equation}
	\begin{equation}
		\label{eq_prob_A_3}
		\Prb[\mathit{BadCom}_2] \leq (n_\text{v} n_{\text{qs}}+  n_{\text{qh}})^2 / 2^{l_\text{hash}+1},
	\end{equation}
	\begin{equation}
		\label{eq_prob_A_4}
		\Prb[\mathit{BadCom}_3] \leq n_\text{v} n_{\text{qs}
		} / 2^{l_\text{hash}}.
	\end{equation}
	
	$\mathit{BadProg}$ happens when $T_\text{app}(\widetilde{\mathit{pk}}\| \widetilde{R}\| \overline{data})$ has already been defined. Every time the forger makes an approval query, it uses different $\mathit{data}_i$ (i.e., $c_i$) where $i\neq i^*$. We do not have any knowledge of how the forger chooses it so that we consider $c_i$ is chosen uniformly at random. 
	$c_i$ has $p_\text{mk}$ possible values. $R_i$ has $2^{l_\text{q}-1}$ possible values. There are $n_\text{q}$ assignments to the table in total. Thus, for each approval query, the probability that $T_\text{app}(\widetilde{\mathit{pk}}\| \widetilde{R}\| \overline{data})$ has already been defined is less than $n_\text{q}/ (p_\text{mk}2^{l_\text{q}-1})$. Considering that there are $n_{\text{qs}}$ approval queries in total, the probability of $\mathit{BadProg}$ is:
	\begin{equation}
		\label{eq_prob_A_5}
		\Prb[\mathit{BadProg}] \leq n_{\text{qs}}n_\text{q}/ (p_\text{mk}2^{l_\text{q}-1}).
	\end{equation}
	
	$\mathit{BadOrder}$ happens when the assignment of $T_\text{agg}(L_\mathit{pk}\| \mathit{pk}_{i^*})$ occurs after that of $T_\text{app}(\widetilde{\mathit{pk}}\| \widetilde{R} \| \overline{data})$. Different with that in MuSig, there are at most $n_{\text{qs}}+ n_{\text{qh}}$ assignments of $\Hashagg(L_\text{pk}\| \mathit{pk}_i)$. The probability of $\mathit{BadOrder}$ is:
	\begin{equation}
		\label{eq_prob_A_6}
		\Prb[\mathit{BadOrder}] \leq (n_{\text{qs}}+ n_{\text{qh}})n_\text{q}/2^{l_\text{hash}}.
	\end{equation}
	
	Similarly, the probability of $\mathit{KeyColl}$, which happens when two sets of public keys have the same aggregated key, is changed:
	\begin{equation}
		\label{eq_prob_A_7}
		\Prb[\mathit{KeyColl}]  \leq (n_{\text{qs}}+ n_{\text{qh}})^2 / 2^{l_\text{hash}}.
	\end{equation}
	
	With Equations~\eqref{eq_prob_A_1},~\eqref{eq_prob_A_2},~\eqref{eq_prob_A_3},~\eqref{eq_prob_A_4},~\eqref{eq_prob_A_5},~\eqref{eq_prob_A_6} and~\eqref{eq_prob_A_7}, we give a lower bound for $\mathit{acc}(\mathds{A})$ as follows:
	\begin{equation}
		\label{eq_prb_A}
		\mathit{acc}(\mathds{A}) 
		\geq \epsilon-2(\frac{p_\text{mk}+1}{p_\text{mk}})\frac{n_{\text{qs}}(n_\text{v}n_{\text{qs}}+n_{\text{qh}})}{2^{l_\text{q}}} - \frac{4 (n_\text{v} n_{\text{qs}} + n_{\text{qh}})^2}{2^{l_\text{hash}}}.
	\end{equation}
	
	Now we can construct an algorithm $\mathds{B}$, which runs $\mathds{A}$ twice and forks the execution of the forger on the answer to the query $\Hashapp(\widetilde{\mathit{pk}}\| \widetilde{R} \| \overline{data} )$. With $\mathds{B}$, the discrete logarithm of $\widetilde{\mathit{pk}}$ can be retrieved. According to the generalized forking lemma~\cite{bellare2006multi}, the accepting probability of $\mathds{B}$ can be calculated as follows:
	\begin{equation}
		\begin{split}
			\mathit{acc}(\mathds{B}) \geq & 
			\mathit{acc}(\mathds{A})\cdot (\frac{\mathit{acc}(\mathds{A})}{n_\text{q}}-\frac{1}{2^{l_\text{hash}}})
			\\
			\geq & \frac{\epsilon^2}{n_{\text{qs}}+n_{\text{qh}}+1} -4(\frac{p_\text{mk}+1}{p_\text{mk}})\frac{n_{\text{qs}}(n_\text{v}n_{\text{qs}}+n_{\text{qh}})}{2^{l_\text{q}}}
			\\  & -  \frac{8(n_\text{v} n_{\text{qs}} + n_{\text{qh}})^2+1}{2^{l_\text{hash}}}.
		\end{split}
	\end{equation}
	
	Algorithm $\mathds{C}$ runs $\mathds{B}$ twice and forks the execution on the answer to $\Hashagg(L_\text{pk}\| \mathit{pk}_i)$. With $\mathds{C}$, the discrete logarithm of $\mathit{pk}_{i^*}$ can be retrieved. We omit the details of the double-forking lemma~\cite{maxwell2019simple} for brevity. The accepting probability of $\mathds{C}$ is bounded as follows:
	\begin{equation}
		\begin{split}
			\mathit{acc}(\mathds{C}) \geq & 
			\mathit{acc}(\mathds{B})\cdot (\frac{\mathit{acc}(\mathds{B})}{n_\text{q}-1}-\frac{1}{2^{l_\text{hash}}})
			\\
			\geq & \frac{\epsilon^4}{(n_{\text{qs}}+n_{\text{qh}}+1)^2(n_{\text{qs}}+n_{\text{qh}})} -8(\frac{p_\text{mk}+1}{p_\text{mk}}) 
			\\& \frac{n_{\text{qs}}(n_\text{v}n_{\text{qs}}+n_{\text{qh}})}{2^{l_\text{q}}} 
			- \frac{16(n_\text{v} n_{\text{qs}} + n_{\text{qh}})^2+3}{2^{l_\text{hash}}}.
		\end{split}
	\end{equation}
	
	Compared with that in MuSig proof, the running time of $\mathds{C}$ changes a bit. The size of the table $T_\text{agg}$ is at most $n_\text{v}(n_{\text{qs}}+n_{\text{qh}})$ in $\mathds{A}$ while others remain the same. Suppose the time of each table assignment is $\mathcal{O}(1)$. We use $T_\text{G\_MUL}$ to denote the point multiplication on $\mathbb{G}$.
	The running time of $\mathds{C}$, $T_\mathds{C}$, is as follows:
	\begin{equation}
		T_{\mathds{C}} = 4T_\text{fg} + 4\mathcal{O}(n_\text{v}(n_{\text{qs}}+n_{\text{qh}})) + 4n_\text{v}T_\text{G\_MUL}.
	\end{equation}
\end{proof} 